\documentclass[twocolumn, trackchanges]{aastex63}

\usepackage{amsmath}
\usepackage{subfigure}

\received{\today}
\submitjournal{PASP}

\shorttitle{ATOCA}
\shortauthors{Darveau-Bernier et al.}

\begin{document}
\title{\textsc{ATOCA: an algorithm to treat order contamination. Application to the NIRISS SOSS mode.}}



\correspondingauthor{Antoine Darveau-Bernier}
\email{antoine.darveau-bernier@umontreal.ca}

\author[0000-0002-7786-0661]{Antoine Darveau-Bernier}
\affiliation{Institut de Recherche sur les Exoplan\`etes (iREx), Universit\'e de Montr\'eal, D\'epartement de Physique, C.P. 6128 Succ. Centre-ville, Montr\'eal, \\ QC H3C 3J7, Canada. }

\author[0000-0003-0475-9375]{Lo\"ic Albert}
\affiliation{Institut de Recherche sur les Exoplan\`etes (iREx), Universit\'e de Montr\'eal, D\'epartement de Physique, C.P. 6128 Succ. Centre-ville, Montr\'eal, \\ QC H3C 3J7, Canada. }

\author[0000-0003-4787-2335]{Geert Jan Talens}
\affiliation{Institut de Recherche sur les Exoplan\`etes (iREx), Universit\'e de Montr\'eal, D\'epartement de Physique, C.P. 6128 Succ. Centre-ville, Montr\'eal, \\ QC H3C 3J7, Canada. }
\affiliation{Department of Astrophysical Sciences, Princeton University, 4 Ivy Lane, Princeton, NJ 08544, USA}

\author[0000-0002-6780-4252]{David Lafreni\`ere}
\affiliation{Institut de Recherche sur les Exoplan\`etes (iREx), Universit\'e de Montr\'eal, D\'epartement de Physique, C.P. 6128 Succ. Centre-ville, Montr\'eal, \\ QC H3C 3J7, Canada. }

\author[0000-0002-3328-1203]{Michael Radica}
\affiliation{Institut de Recherche sur les Exoplan\`etes (iREx), Universit\'e de Montr\'eal, D\'epartement de Physique, C.P. 6128 Succ. Centre-ville, Montr\'eal, \\ QC H3C 3J7, Canada. }

\author[0000-0001-5485-4675]{Ren\'e Doyon}
\affiliation{Institut de Recherche sur les Exoplan\`etes (iREx), Universit\'e de Montr\'eal, D\'epartement de Physique, C.P. 6128 Succ. Centre-ville, Montr\'eal, \\ QC H3C 3J7, Canada. }

\author[0000-0003-4166-4121]{Neil~J.~Cook}
\affiliation{Institut de Recherche sur les Exoplan\`etes (iREx), Universit\'e de Montr\'eal, D\'epartement de Physique, C.P. 6128 Succ. Centre-ville, Montr\'eal, \\ QC H3C 3J7, Canada. }

\author[0000-0002-5904-1865]{Jason F. Rowe}
\affiliation{Bishop's University, 2600 College Street, Sherbrooke, QC J1M 1Z7, Canada \\}
\affiliation{Institut de Recherche sur les Exoplan\`etes (iREx), Universit\'e de Montr\'eal, D\'epartement de Physique, C.P. 6128 Succ. Centre-ville, Montr\'eal, \\ QC H3C 3J7, Canada. }


\author[0000-0003-3506-5667]{\'Etienne Artigau} 
\affiliation{Institut de Recherche sur les Exoplan\`etes (iREx), Universit\'e de Montr\'eal, D\'epartement de Physique, C.P. 6128 Succ. Centre-ville, Montr\'eal, \\ QC H3C 3J7, Canada. }

\author[0000-0001-5578-1498]{Bj\"orn Benneke}
\affiliation{Institut de Recherche sur les Exoplan\`etes (iREx), Universit\'e de Montr\'eal, D\'epartement de Physique, C.P. 6128 Succ. Centre-ville, Montr\'eal, \\ QC H3C 3J7, Canada. }

\author[0000-0001-6129-5699]{Nicolas Cowan}
\affiliation{Department of Earth \& Planetary Sciences, McGill University, 3450 rue University, Montréal, QC H3A 0E8, Canada\\}
\affiliation{Department of Physics, McGill University, 3600 rue University, Montréal, QC H3A 2T8, Canada\\}

\author[0000-0003-4987-6591]{Lisa Dang}
\affiliation{Department of Physics, McGill University, 3600 rue University, Montréal, QC H3A 2T8, Canada\\}
\affiliation{Institut de Recherche sur les Exoplan\`etes (iREx), Universit\'e de Montr\'eal, D\'epartement de Physique, C.P. 6128 Succ. Centre-ville, Montr\'eal, \\ QC H3C 3J7, Canada. }

\author[0000-0001-9513-1449]{N\'estor Espinoza}
\affiliation{Space Telescope Science Institute, 3700 San Martin Drive, Baltimore, MD 21218, USA}
\affiliation{Department of Physics \& Astronomy, Johns Hopkins University, 3400 N Charles St, Baltimore, MD 21218, USA}

\author[0000-0002-6773-459X]{Doug Johnstone}
\affiliation{NRC Herzberg Astronomy and Astrophysics, 5071 West Saanich Rd, Victoria, BC, V9E 2E7, Canada}
\affiliation{Department of Physics and Astronomy, University of Victoria, Victoria, BC, V8P 5C2, Canada}

\author[0000-0002-0436-1802]{Lisa Kaltenegger}
\affiliation{Department of Astronomy and Carl Sagan Institute, Cornell University, 302 Space Sciences Building, Ithaca, NY 14853, USA}

\author[0000-0003-4676-0622]{Olivia Lim}
\affiliation{Institut de Recherche sur les Exoplan\`etes (iREx), Universit\'e de Montr\'eal, D\'epartement de Physique, C.P. 6128 Succ. Centre-ville, Montr\'eal, \\ QC H3C 3J7, Canada. }

\author[0000-0002-8573-805X]{Stefan Pelletier}
\affiliation{Institut de Recherche sur les Exoplan\`etes (iREx), Universit\'e de Montr\'eal, D\'epartement de Physique, C.P. 6128 Succ. Centre-ville, Montr\'eal, \\ QC H3C 3J7, Canada. }

\author[0000-0002-2875-917X]{Caroline Piaulet}
\affiliation{Institut de Recherche sur les Exoplan\`etes (iREx), Universit\'e de Montr\'eal, D\'epartement de Physique, C.P. 6128 Succ. Centre-ville, Montr\'eal, \\ QC H3C 3J7, Canada. }

\author[0000-0001-8127-5775]{Arpita Roy}
\affiliation{Space Telescope Science Institute, 3700 San Martin Drive, Baltimore, MD 21218, USA}
\affiliation{Department of Physics \& Astronomy, Johns Hopkins University, 3400 N Charles St, Baltimore, MD 21218, USA}

\author[0000-0001-6809-3520]{Pierre-Alexis Roy}
\affiliation{Institut de Recherche sur les Exoplan\`etes (iREx), Universit\'e de Montr\'eal, D\'epartement de Physique, C.P. 6128 Succ. Centre-ville, Montr\'eal, \\ QC H3C 3J7, Canada. }

\author[0000-0001-9987-467X]{Jared Splinter}
\affiliation{Department of Earth \& Planetary Sciences, McGill University, 3450 rue University, Montréal, QC H3A 0E8, Canada\\}

\author[0000-0003-4844-9838]{Jake Taylor}
\affiliation{Institut de Recherche sur les Exoplan\`etes (iREx), Universit\'e de Montr\'eal, D\'epartement de Physique, C.P. 6128 Succ. Centre-ville, Montr\'eal, \\ QC H3C 3J7, Canada. }
s
\author[0000-0001-7836-1787]{Jake D. Turner}
\affiliation{Department of Astronomy and Carl Sagan Institute, Cornell University, Ithaca, NY 14853, USA}

%
%
%
%

\begin{abstract}
After a successful launch, the James Webb Space Telescope is preparing to undertake one of its principal missions, the characterization of the atmospheres of exoplanets. The Single Object Slitless Spectroscopy (SOSS) mode of the Near Infrared Imager and Slitless Spectrograph (NIRISS) is the only observing mode that has been specifically designed for this objective. It features a wide simultaneous spectral range (0.6--2.8\,\micron) through two spectral diffraction orders. However, due to mechanical constraints, these two orders overlap slightly over a short range, potentially introducing a ``contamination'' signal in the extracted spectrum.
We show that for a typical box extraction, this contaminating signal amounts to 1\% or less over the 1.6--2.8\,\micron\ range (order 1), and up to 1\% over the 0.85--0.95\,\micron\ range (order 2). For observations of exoplanet atmospheres (transits, eclipses or phase curves) where only temporal variations in flux matter, the contamination signal typically biases the results by order of 1\% of the planetary atmosphere spectral features strength. To address this problem, we developed the Algorithm to Treat Order ContAmination (ATOCA). By constructing a linear model of each pixel on the detector, treating the underlying incident spectrum as a free variable, ATOCA is able to perform a simultaneous extraction of both orders. We show that, given appropriate estimates of the spatial trace profiles, the throughputs, the wavelength solutions, as well as the spectral resolution kernels for each order, it is possible to obtain an extracted spectrum accurate to within 10\,ppm over the full spectral range.
\\
\end{abstract}


\section{Introduction}
\label{sec:Intro}

\begin{figure*}[tbph]
    \centering
    \includegraphics[width=\linewidth]{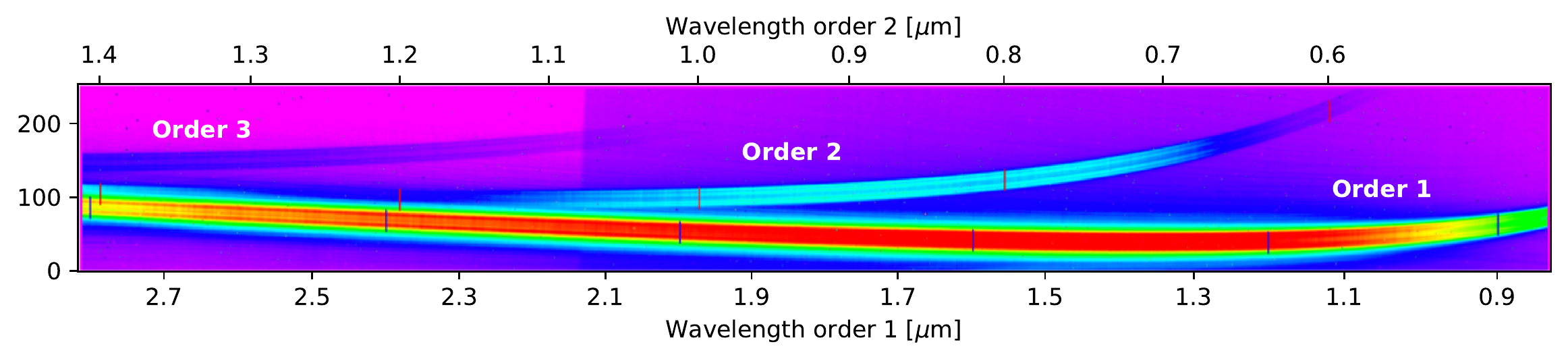}
    \caption{High signal-to-noise stack of a SOSS mode observation on a tungsten lamp obtained at cryogenic vacuum testing (CV3) of the telescope and instruments. Order 1 is the most apparent feature extending from 0.83\,\micron\ on the right to 2.8\,\micron\ on the left portion of the image.  The second order can be seen starting at 0.6\,\micron\ at the top right part of the image and extends out to 1.4\,\micron.  At approximately 1.1\,\micron\ the second order is significantly blended with the first order.  The faint third order can be observed above the 2nd order. The overlap between the spectral orders 1 and 2 on the left side of the image complicates the spectrum extraction and motivates this paper. Since the order 2 covers shorter wavelengths than order 1, this problem should be even more striking in actual astrophysical targets which are warmer ($T\ge3000$\,K) than the tungsten filament used in the laboratory (1500\,K). The vertical lines were added to mark the wavelengths, from right to left, at 0.9, 1.2, 1.6, 2.0, 2.4, 2.8\,$\mu$m for order 1 (in black) and at 0.6, 0.8, 1.0, 1.2, 1.4\,$\mu$m for order 2 (in red).}
    \label{fig:sossmode}
\end{figure*}

One of the key observing modes of the Near Infrared and Slitless Spectrograph (NIRISS, Doyon et al., in prep) onboard the James Webb Space Telescope (JWST) is the Single Object Slitless Spectroscopy (SOSS) mode (Albert et al., in prep). It enables time-series spectroscopy in the 0.6--2.8\,\micron\ range for bright targets, which is of particular use for exoplanet transit spectroscopy. Indeed, simulations have demonstrated SOSS as a key mode to use with JWST on the brightest exoplanet targets \citep{greene.2016,batalha.2017,louie.2018,schlawin.2018} and it has been selected by multiple Cycle 1 programs, including the Early Release Science program DD-ERS 1366 \citep{1366jwst, bean.2018}, as well as many Guaranteed Time Observations, e.g., GTO 1201 \citep{1201jwst} and the General Observer Programs 1935 \citep{1935jwst}, 2062 \citep{2062jwst}, 2113 \citep{2113jwst}, 2589 \citep{2589jwst}, 2594 \citep{2594jwst} and 2722 \citep{2722jwst}. SOSS uses the GR700XD cross-dispersion grating prism (Doyon et al., in prep) in the pupil wheel of NIRISS to produce a series of three spectral traces: order 1 (0.83--2.8\,\micron), order 2 (0.6--1.4\,\micron) and order 3 (0.6--0.95\,\micron). In practice, order 3 does not warrant much consideration due to its faint signal and the fact that it does not increase the wavelength domain. A slight (22 pixel wide) defocus along the spatial axis is purposely included to enable observations of bright targets without saturating the detector pixels. Mechanical constraints in the thickness of the GR700XD element at the design phase prevented the first and second orders from being fully separated, resulting in an overlap by about half the trace width towards the red wavelength ends of the traces (See Figure~\ref{fig:sossmode})

As a result, established methods for spectrum extraction cannot be applied directly to the regions affected by contamination. Typically, at high signal-to-noise, a box-extraction method \citep{deboer.1981} is preferred, which is performed by simply summing over the spatial axis all pixels located within a fixed-width aperture. This technique has been utilized in many space-based relative spectral measurements of exoplanets to date, for example: transits, eclipses or phase curves \citep[e.g.,][]{deming.2013, wakeford.2013, sing.2015, evans.2017} as well as in many ground-based observations \citep[e.g.,][]{jordan.2013, diamond.2018}. The main advantages are the fact that a box extraction is easy to implement and that it is less prone to modelling errors.
On the other hand, at lower signal-to-noise, an optimal extraction \citep{horne.1986, robertson.1986, marsh.1989} is often better. Indeed, by weighting the pixels according to their relative contribution to the signal, a better precision can be reached. This requires the determination of a spatial profile, which is a delicate task that can introduce biases in the resulting spectra  \citep{horne.1986, jordan.2013}. Nevertheless, it is still used in the exoplanet community to perform spectrophotometric measurements from space \citep[e.g.,][]{kreidberg.2014, stevenson.2019} or from the ground \citep[e.g.,][]{berta.2011, stevenson.2014}, with comparable results. However, these two methods have no mechanism to distinguish between contributions from overlapping traces from different sources or diffraction orders.

Yet, the challenge of extracting spectra from blended sources is not unprecedented. It was needed notably in the context of long-slit spectroscopy for several science applications, such as the observation of galactic nuclei \citep{lucy.2003} or crowded star fields \citep{hynes.2002}. In fact, a task as common as a simple sky subtraction is by itself a type of decontamination. Hence, various techniques have been proposed over the past two decades \citep[e.g.,][]{hynes.2002, khmil.2002, lucy.2003, bolton.2010}. However, due to the particularity of the NIRISS SOSS mode and the precision it requires, it was necessary to develop a dedicated algorithm.

In this article, we present ATOCA; an algorithm designed to properly decontaminate and extract overlapping orders. Though the methods that make up the ATOCA algorithm can be applied generally to the problem of extracting overlapping spectral orders, we focus here on the NIRISS/SOSS mode of JWST. Proper extraction of SOSS observations was our primary motivation for creating ATOCA, and the algorithm has been made part of the official JWST pipeline\footnote{\url{https://github.com/spacetelescope/jwst}} -- the data management system (DMS) -- as part of the stage 2 spectral extraction step. 




The article is divided as follow: Section \ref{sec:the_problem} presents an estimate of the level of contamination that is expected with the NIRISS/SOSS mode. Then, the algorithm and its implementation are presented in sections \ref{sec:algorithm} and \ref{sec:implementation}, followed by section \ref{sec:validation} where we evaluate the performance of the decontamination.

\section{The SOSS Trace Overlap Problem} \label{sec:the_problem}


\begin{figure*}
\centering
    \begin{subfigure}
    \centering
    \includegraphics[scale=0.67]{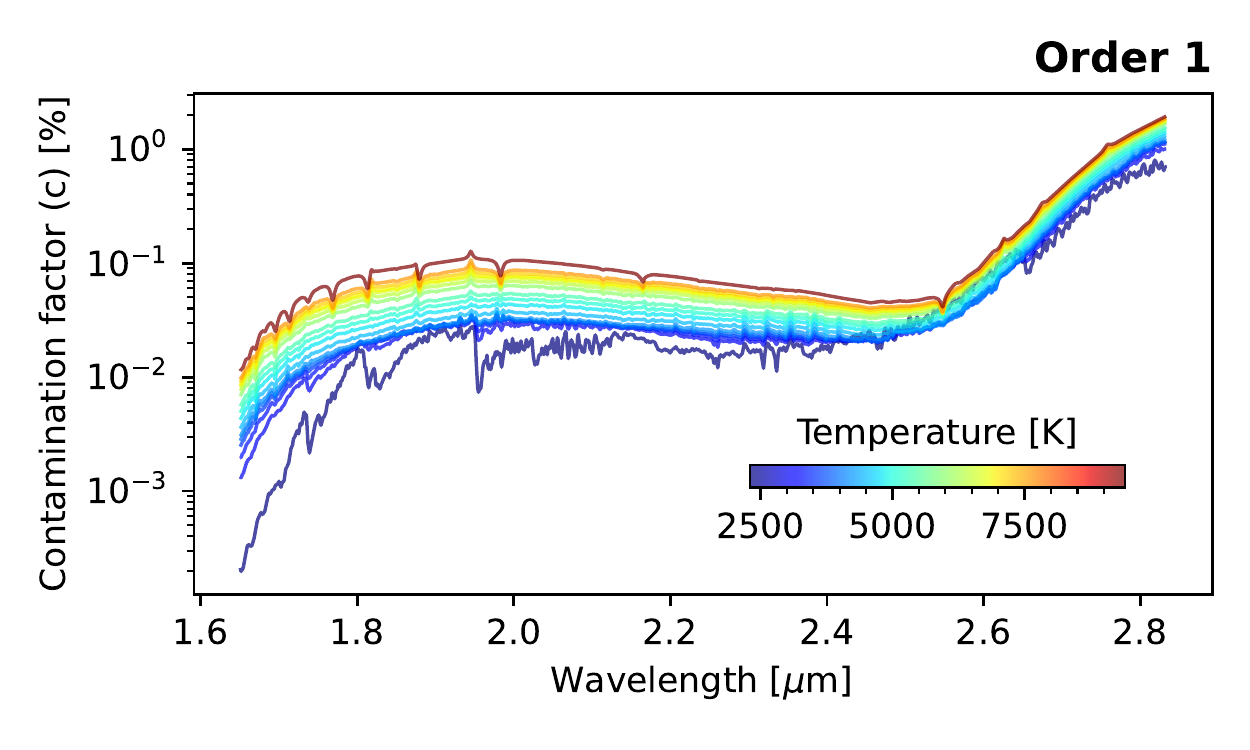} %
    \end{subfigure}
    \begin{subfigure}
    \centering
    \includegraphics[scale=0.67]{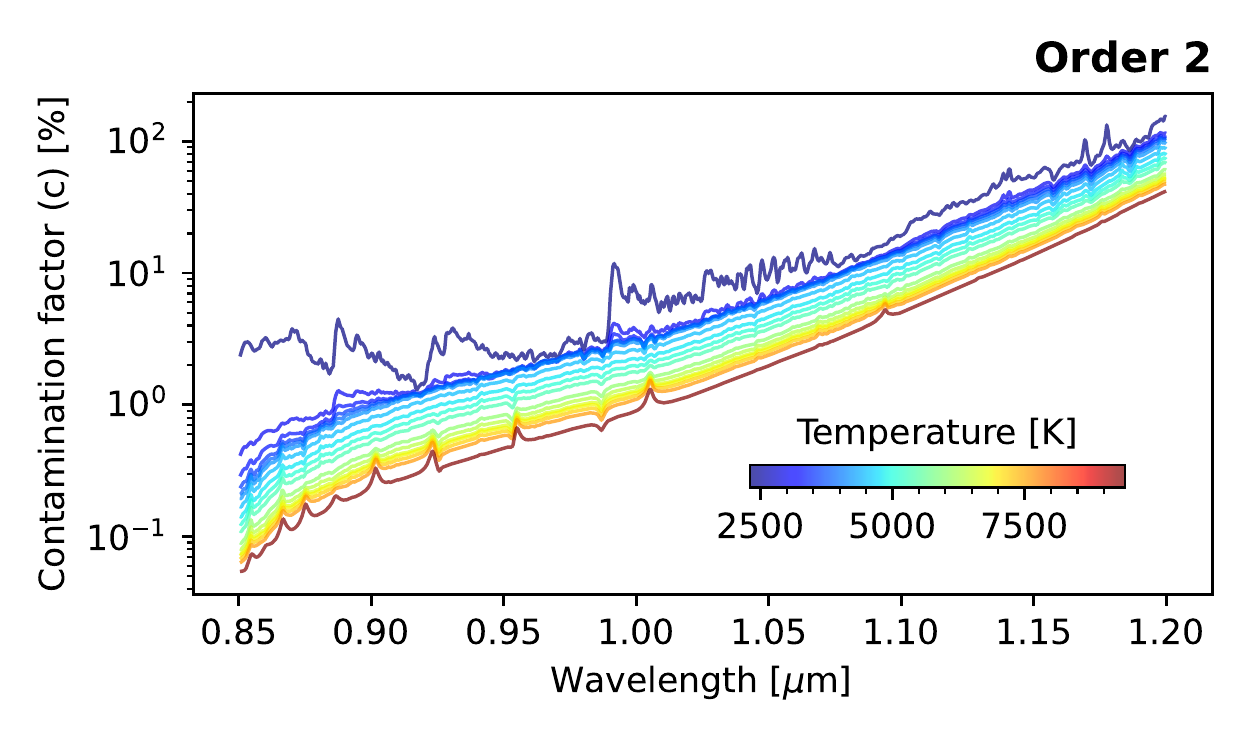} %
    \end{subfigure}
    \caption{Contamination factors (see equation \ref{eq:transit_contamination}) for a range of stellar effective temperatures (color-coded). These values hold for a standard box extraction, using a 25-pixel aperture. The wavelength domain is not entirely shown here; the first order (left panel) is virtually uncontaminated below 1.6\,\micron\ whereas the second order (right panel) contamination levels increase exponentially at longer wavelengths ($>1.1\,\micron$).}%
    \label{fig:contamination_factors}
\end{figure*}

\begin{figure*}
\centering
    \begin{subfigure}
    \centering
    \includegraphics[scale=0.67]{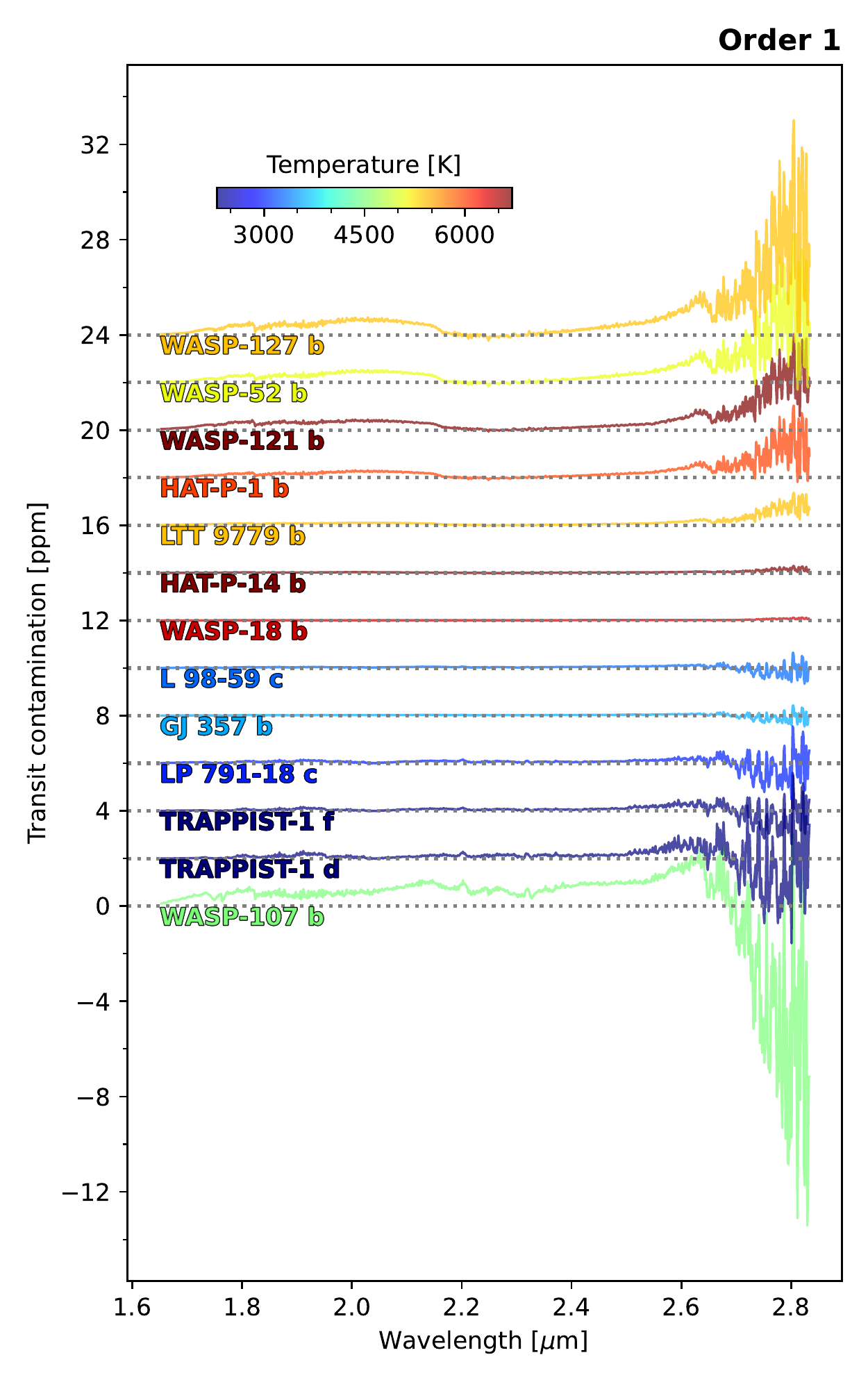}
    \end{subfigure}
    \begin{subfigure}
    \centering
    \includegraphics[scale=0.67]{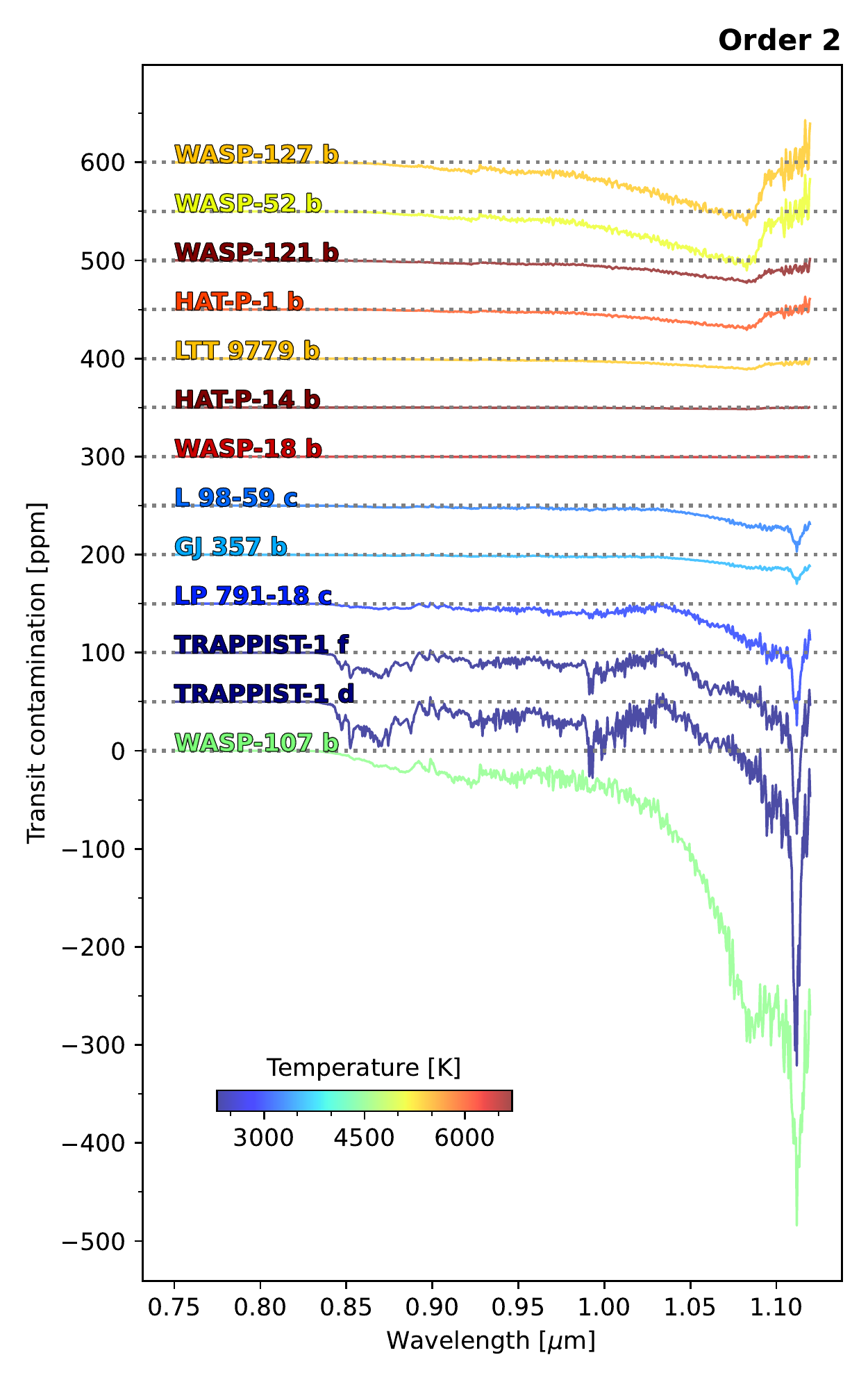}
    \end{subfigure}
    \caption{Expected contamination levels during transit for a diverse sample of exoplanets. Each target is shifted vertically by 2 ppm for the first order (left panel) and by 50 ppm for the second order (right panel). A horizontal dotted line is drawn for each model to mark its zero point. The color of the lines refers to the effective temperature of the host star, which influences the level of contamination. As in Figure \ref{fig:contamination_factors}, these estimates were made assuming a 25-pixel-wide box extraction and the wavelength domain was limited for the same reasons. }%
    \label{fig:transit_contamination}
\end{figure*}


The optics of the SOSS mode were designed with strong mechanical constraints, one of which, the total thickness of the GR700XD element, prevented the cross-dispersing prism from being sufficiently inclined to cleanly separate spectral orders 1 and 2 (see Albert et al. in prep). As a result, the red end of order 2 ($\lambda \geq 1.1\mu$m) partially overlaps with that of order 1 ($\lambda \geq 2.2\mu$m)  (see Figure \ref{fig:sossmode}). This cross contamination of the signals is a major issue during spectrum extraction and will bias results using the simple aperture-based methods discussed in Section \ref{sec:Intro}.
The amount of contamination can be characterized by measuring the contaminating signal present in the extraction region, $F_\mathrm{contam}$, allowing for the definition of a contamination factor, $c_{order}$, via the following ratio:
\begin{equation} \label{eq:contamination_factor}
    c_{order} = \frac{F_{\mathrm{contam}}}{F_{order}} \,.
\end{equation}
Here, $F_{order}$ is the flux that would be extracted from the targeted order if there was no contamination. For example, for the first order, $F_\mathrm{contam}$ is given by the signal from the second order that is overlapping with the region of interest. Based on laboratory measurements of the blaze function, detector efficiency, and end-to-end throughput of the telescope and NIRISS instrument, it is possible to simulate this contamination factor for any given extraction method. 

Figure \ref{fig:contamination_factors} shows the contamination factors, $c$, for a standard box extraction using a 25-pixel aperture. The  simulations were made using ATOCA and are described in Section \ref{sec:simulation}. The aperture width was determined by minimizing the dispersion of the white light-curve over the contaminated band-pass (2.0--2.8\,\micron). To get a realistic estimation of the dispersion, we had to consider the expected jitter on the telescope pointing $\sim$ 5\,mas (see Albert et al., in prep.) which has the effect to increase the width of the required aperture. The incident fluxes at different stellar effective temperatures were modeled using PHOENIX HiRes synthetic spectra \citep{husser.2013}. Note also that the contamination factor is presented in \% to make a distinction with the transit depth (which is is typically in ppm), both being relative quantities.

For the first diffraction order, most of the affected wavelength range exhibits levels of contamination below 0.1\,\%, until 2.6\,\micron, after which $c_1$ increases exponentially up to 2\,\%. The stellar effective temperature also has a significant effect, since the second order covers shorter wavelengths ($0.6\leq\lambda\leq1.4\,\mu$m) than the first order ($0.83\leq\lambda\leq2.8\,\mu$m), hence a star with a stronger relative flux contribution at short wavelengths will be more affected. The second order, on the other hand, suffers more drastically from contamination, reaching levels above 100\%. Fortunately, the wavelength domain affected ($\lambda\geq0.85\,\mu$m) is shared with order 1 (and is within the region where order 1 does not suffer from contamination), so very little information is lost. The wavelengths towards the blue end of the spectrum ($\lambda\leq0.85\,\mu$m) and complementary to order 1 correspond to the part of the detector where the orders' spatial positions deviate from one another, creating a drop in the contamination levels.

The above discussion holds for any absolute flux measurement, but what is the impact for the intended application of the SOSS mode --- exoplanet time-series --- whose measurements are relative?

The fluxes measured in order 1 will be a combination of the true flux in that order, $F_1$, contaminated by some flux from order 2, $F_\mathrm{contam}$. Assuming a simplified top-hat model for an exoplanet transit (no limb darkening, instantaneous ingress and egress, non-grazing) this flux can be calculated, for cases in and out of transit, via:

\begin{equation}
    F_{out} = F_1 + F_\mathrm{contam}
\end{equation}

\begin{equation}
    F_{in} =  (1-d-\delta_1(\lambda_1)) F_1 + (1-d-\delta_2(\lambda_2)) F_\mathrm{contam}
\end{equation}

where $F_{out}$ and $F_{in}$ are respectively the mean flux outside of transit and during transit measured by an extraction around the first order's trace, $d$ is the transit depth due to the opaque planet (i.e., without considering an atmosphere) 
and $\delta_1(\lambda_1)$ and $\delta_2(\lambda_2)$ are the wavelength-dependent transit depths due to the planet's atmosphere for orders 1 and 2, respectively. $\lambda_1$ and $\lambda_2$ are the wavelength solutions at each order, which are both a function of the column position, $x$, such that $\lambda_1(x)$ and $\lambda_2(x)$. 
The transit depth measured on a contaminated trace is, by definition:

\begin{equation}
    D = 1 - F_{in}/F_{out}. 
\end{equation}

Recalling the order contamination factor from equation \ref{eq:contamination_factor}, then the transit depth can be written:

\begin{equation} \label{eq:transit_contamination}
    D = d + \delta_1(\lambda_1) +  \frac{c_1}{1+c_1} \left( \delta_2(\lambda_2) - \delta_1(\lambda_1) \right).
\end{equation}

In the case where there is no chromatic variation in the atmospheric signal (i.e., a flat transmission spectrum), $\delta_2(\lambda_2) = \delta_1(\lambda_1)$ so $D = d + \delta_1(\lambda_1)$. Therefore, contamination has no bearing on the retrieved transit depth. In other words, the second order contaminating signal changes by exactly the same relative amount during transit as the first order signal.

In the case where the atmospheric signal is different at the two overlapping wavelengths, then the difference

\begin{equation}
    \Delta(x) = \delta_2(\lambda_2(x)) - \delta_1(\lambda_1(x))   
\end{equation}

modulated by $c_1/(1 + c_1)$ will affect the transit depth (i.e., the last term in equation \ref{eq:transit_contamination}). To make a distinction with the contamination factor $c$, we will use the name ``transit contamination'' to refer to this last term of equation \ref{eq:transit_contamination}. Generally, $\Delta$ will be about the same order of magnitude as $\delta_1$ and $\delta_2$, so a good estimation can be drawn simply from the contamination term. Moreover, for order 1, $c$ will be small, hence $c/(c+1) \approx c$. So, concerning the first order's relative measurement, the spurious signal can be approximated to less than 1\% (i.e., $c_1$) of the chromatic contribution of the transit signal, $\delta_1$. For example, if we take a hypothetical transmission spectrum with a spectral feature for the first order of $\delta_1(2.7\micron) = 300$\,ppm above the mean transit depth $d$. Let's also assume that there is a spectral feature from the second order at the corresponding columns (see Figure \ref{fig:sossmode}) of $\delta_2(1.35\micron) = -200$\,ppm, i.e., $200$\,ppm below the mean transit depth. This would result in a difference of $\Delta(x)=500$\,ppm and the resulting contamination signal will be around 1.5\,ppm, considering a contamination factor of $c_1\approx3$\,\% (see Figure \ref{fig:contamination_factors}).
%



Nevertheless, to fully grasp the importance of this effect, we computed the resulting contamination signal in transmission for a variety of exoplanets, most of them being part of the NIRISS Exploration of the Atmospheric diversity of Transiting exoplanets (NEAT) GTO program \citep{1201jwst}. The results are shown in Figure \ref{fig:transit_contamination}. The transit models for each planet were produced using the SCARLET atmosphere framework \citep{benneke_atmospheric_2012, benneke_how_2013, benneke.2015} assuming, for simplicity, cloud-free atmospheres with solar elemental abundances and chemical equilibrium. These assumptions generally lead to stronger signals, hence upper limits on the estimates. For the first order, the transit contamination is constrained below 8\,ppm for all targets except WASP-107\,b, for which it reaches almost $-12$\,ppm. This corresponds to $\approx$1\% of the planet atmospheric signal, as expected. For example, the transmission spectrum of WASP-107\,b presents spectroscopic variations around 2500\,ppm  (see Figure \ref{fig:hot_jup_tr}), which would lead to an expected transit contamination signal of 25\,ppm, not far from 12\,ppm value. On the other hand, the second order is much more affected, with levels around 100\,ppm or more in the longer wavelength range shown in Figure \ref{fig:transit_contamination}. This contamination comes from the wings of the first order's spatial profile. The longer wavelengths are not presented since the second order becomes almost completely diluted into the first order (see Figure \ref{fig:sossmode}). On the other side, below  0.8\,\micron, where the second order contributes unique wavelength coverage, the transit contamination seems to vanish. However, this drastic drop in contamination is attributable to the simulations, as discussed in Section \ref{sec:results}.

Whilst the systematic error on the transit signal may seems small, it must not be taken lightly and should be prevented using the ATOCA extraction method presented in the next section. We also want to emphasize that this is a systematic error, and not a randomly distributed source of noise such as shot noise. These estimations could also be worsened by any other relative signals that depends on wavelength; like limb darkening and stellar contamination from unocculted regions \citep[e.g.][]{rackham.2018, genest.2022}. Moreover, the examples presented here assume that the trace shape is perfectly stable within a whole time-series and that the trace position is varying within the expectation, i.e., following a random normal distribution with a dispersion of 5\,mas. In the context of real observations, these assumptions may not be true due to the finite pointing precision of the Fine Guidance Sensor (FGS) and possible variations in the point spread function (breathing effects, wavefront variations, etc.). Therefore, to obtain the most stable transmission spectrum, one may need to increase the width of the aperture (along the spatial axis) used for the box extraction \citep[e.g.,][]{diamond.2018, mikal-evans.2021} in order to minimize the variation due to the signal moving in and out of the aperture, hence increase the contribution of the contaminating order. Furthermore, even in the context of standard extractions, ATOCA will help to calibrate and extract the one-dimensional spectra. The contamination would also need to be properly characterized by identifying the contribution of each order to ensure the reliability of any results. Finally, extraction using ATOCA will ensure that science applications needing absolute flux calibration can be realized with SOSS.

\section{Description of the Extraction Method} \label{sec:algorithm}

\begin{figure*}
\centering
\includegraphics[scale=0.8]{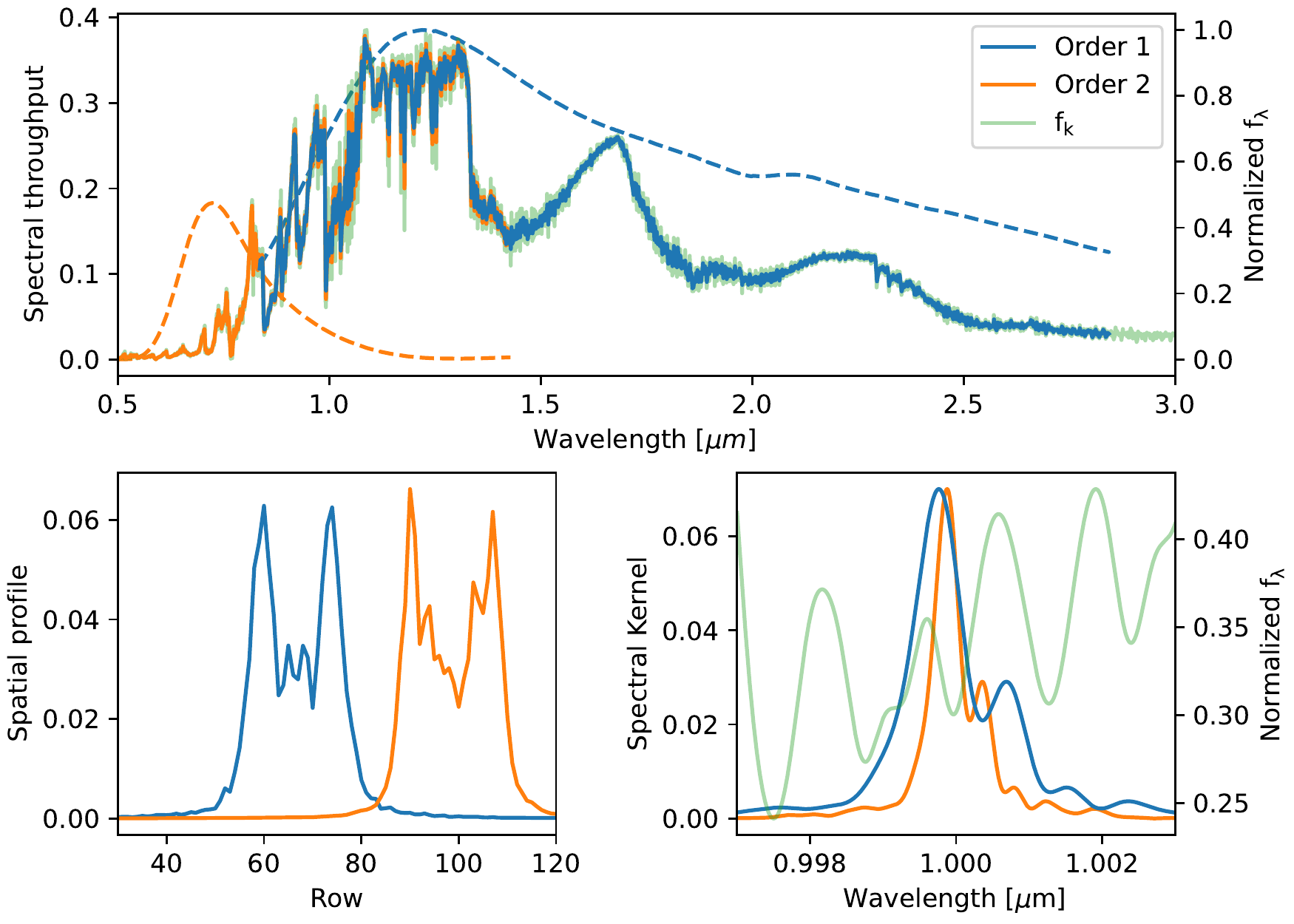}
\caption{\label{fig:2}
Spectrum and associated spectral profiles for a 2300\,K star, along with contributions from the first (blue) and second (orange) orders. 
The top panel shows the spectral throughputs (dashed lines) and the flux density of a 2300\,K PHOENIX spectrum downgraded at each order's resolution and at a higher resolution representative of the underlying flux (in green). The bottom left panel displays the spatial profiles along a given column where the overlap occurs. At the bottom right, an example of the convolution kernels centered at $1 \, \mu \rm m$ is shown. The green curve is the underlying flux (same as top panel). }

\end{figure*}

The core idea behind ATOCA is to determine the underlying flux by fitting each order directly and simultaneously on the detector image, pixel by pixel. To do so, we first need to establish a linear model of each individual pixel with the flux as independent variable. More precisely, we need to discretize the flux by evaluating it on a wavelength grid. Each of the nodes (or elements) of the resulting flux array is an independent variable. Then, by minimizing the $\chi^2$ with respect to each of these nodes, we are able to express the flux as the solution of a linear system, which ultimately enables us to explicitly extract it. Hence, no forward modelling of the flux is needed for an extraction. However, to be accurate, it requires a thorough knowledge of the detector's properties. The formalism of the algorithm is described below. 

For the following equations, each pixel that will be used for the fit will be labeled by the index $i$ and the total number of relevant pixels is given by $N_i$. There is no need to account for the two-dimensional nature of the detector with additional indices. Each diffraction order also needs to be identified, here using the index $n$. Now, to determine the flux falling on each pixel, we need for each order: 1) the wavelength solution, 2) the spectral throughput \footnote{Throughput is defined here as an end-to-end wavelength-dependent transmission (detected flux divided by the flux impinging the telescope).}, and 3) the spatial throughput. From the wavelength solution, we can define the central wavelength of a pixel $i$ at order $n$ as $\lambda_{ni}$. It will also be useful to define $\lambda^+_{ni}$ and $\lambda^-_{ni}$; the wavelengths at the pixel borders in the spectral direction, and $\Delta \lambda_{ni}=\lambda^+_{ni} - \lambda^-_{ni}$; the pixel spectral coverage. The spectral and spatial throughputs are given respectively by the function $T_n(\lambda)$ and the constant $P_{ni}$. Take notice that the former depends on the wavelength but not explicitly on the pixel, whereas the opposite is true for the latter. Finally, let the spectral flux density of the target incident upon the spectrograph be $f(\lambda)$. 

It is also important to consider that the flux density is seen by each order at a different resolution, so for an order $n$, we need an additional input: 4) the spectral resolution kernel $\kappa(\tilde{\lambda}, \lambda)$. It specifies the convolved flux $\tilde{f}$ in relation with an incident flux through the equation

\begin{equation} \label{eq:kernel}
    \tilde{f}_n(\tilde{\lambda}) = \int_{0}^{\infty} \kappa_n(\tilde{\lambda}, \lambda) f(\lambda) d\lambda \, .
\end{equation}

Figure \ref{fig:2} presents some visualizations of the aforementioned quantities. The difference between the two orders after convolution with the resolution kernels becomes apparent in the overlapping wavelength range, where the first order (blue curve) is not superimposed perfectly on the second order (orange curve).

\subsection{The model}

With all this in place, we are now able to define a model of the detector. The number of photo-electrons detected by pixel $i$ can be represented, up to a multiplicative constant, as
\begin{equation}\label{eq:pix_model}
\begin{aligned}
    M_{i} & = \sum_n \int_{\lambda_{ni}^-}^{\lambda_{ni}^+} P_{ni}T_n(\lambda)\tilde{f}_n(\lambda)\lambda d\lambda \\
& = \sum_n \int_{\lambda_{ni}^-}^{\lambda_{ni}^+} a_{ni}(\lambda)\tilde{f}_n(\lambda)d\lambda,
\end{aligned}
\end{equation}
where $a_{ni}(\lambda)$ accounts for all the coefficients that are not the flux. Note that the summation is made over all orders $n$ that contribute to the signal measured by a given pixel $i$. In the case of the NIRISS/SOSS mode, the index $n$ covers only the first and second orders; the third order is not considered since it does not cover the same pixels. To translate this model into a numerical form, we can define a grid where $f$ is projected, labeled by the index $k$ so that $f(\lambda_k) = f_k$ and $\Delta \lambda_k = \lambda_{k+1} - \lambda_k$. The length of the discretized grid would then be given by $N_k$, so that $1 \leq k \leq N_k$.  Similarly, $N_{\tilde{k}}$ is the length of the convolved flux $\tilde{f}$, so that $1 \leq \tilde{k} \leq N_{\tilde{k}}$. We also need a numerical form of this integral. There are multiple ways to do this, but for ATOCA, we use the trapezoidal method on a specified grid as illustrated in Figure \ref{fig:trpz}. The details of this method are in Section \ref{sec:trpz}. Independently of the chosen integration technique, the numerical form of the integral will look like,
\begin{equation}\label{eq:pix_model_num}
    M_i = \sum_n \sum_{\tilde{k}} w_{in\tilde{k}} a_{in\tilde{k}}\tilde{f}_{n\tilde{k}},
\end{equation}
with $w_{in\tilde{k}}$ given by the integration method. To link the diffraction orders, we want to write these equations according to the underlying flux $f(\lambda)$, following equation \ref{eq:kernel}. In the numerical form,
\begin{equation}
    \tilde{f}_{n\tilde{k}} = \sum_k \kappa_{n\tilde{k}k}f_{k}
\end{equation}
with $\kappa_{n\tilde{k}k}$ being the coefficients of the convolution kernel at order $n$. The numerical integration method as well as the kernel are comprised in them.

Finally, we have that
\begin{equation}
    M_i = \sum_n \sum_{\tilde{k}} w_{in\tilde{k}} \, a_{in\tilde{k}}
          \sum_k \kappa_{n\tilde{k}k}f_k \, .
\end{equation}
This equation can be written in a more intuitive matrix form as
\begin{equation}
    \bigg(M\bigg)_{N_i}
        = \sum_n \bigg(w_n a_n\bigg)_{N_i \times N_{\tilde{k}}}
                 \bigg(\kappa_n\bigg)_{N_{\tilde{k}} \times N_k}
                 \bigg(f\bigg)_{N_k} \, .
\end{equation}
To simplify the notation again, we can put all the coefficients for each order in single matrices $\textbf{b}_n$ with dimensions $N_i \times N_k$,
\begin{equation} \label{eq:bn}
    \bigg(M\bigg)_{N_i}
        = \sum_n \bigg(b_n\bigg)_{N_i \times N_k}
                 \bigg(f\bigg)_{N_k} \, ,
\end{equation}
and add them together in one matrix $\textbf{B}$ to have a final model of each valid pixel given by
\begin{equation}\label{eq:B}
    \boxed{
        \textbf{M}_{N_i}
            = \textbf{B}_{N_i \times N_k} \textbf{f}_{N_k}
    } \, .
\end{equation}
This result is one of the main utilities of ATOCA, which is a linear model of the full NIRISS/SOSS detector. One could use it to generate quick simulations, given a model of the incident flux.

\subsection{Solving for \textbf{f}}
Now that we have a model of the intensity at each relevant pixel, we can link their individual measured intensity $D_i$, with $f_i$ by fitting directly the pixel model on the detector using a $\chi^2$ minimization. Given the following equation,
\begin{equation}\label{eq:chi2}
    \chi^2 = \left\| \frac{\textbf{D} - \textbf{M}}{\mathbf{\sigma}} \right\|^2 \, .
\end{equation}
with \textbf{D} being the array of measured intensities on each pixels $i$ and $\mathbf{\sigma}$ the array of their uncertainties, then the best solution can be found by imposing
\begin{equation}
    \frac{d \chi^2}{d f_k} = 0 \, .
\end{equation}
The detailed calculations found in the Appendix \ref{ssec:chi2} lead to the following system of equations:
\begin{equation}\label{eq:solution}
    \left(\frac{\textbf{B}}{\mathbf{\sigma}}\right)^T_{\rm N_k \times N_i}  \left(\frac{\textbf{D}}{\mathbf{\sigma}}\right)_{\rm N_i}  
    = \left(\frac{\textbf{B}}{\mathbf{\sigma}}\right)_{\rm N_k \times N_i}^T
      \left(\frac{\textbf{B}}{\mathbf{\sigma}}\right)_{\rm N_i \times N_k}
      \mathbf{f}_{\rm N_k} \, ,
\end{equation}
where
\begin{equation}
    \left(\frac{\textbf{B}}{\mathbf{\sigma}}\right)_{\rm N_i \times N_k} = \mathrm{diag} \left(\frac{1}{\mathbf{\sigma}}\right)_{\rm N_i \times N_i} \textbf{B}_{\rm N_i \times N_k} \, .
\end{equation}
This system can now be solved for $\mathbf{f}$. A comparison with the optimal extraction method \citep{horne.1986} is presented in Appendix \ref{sec:comparison}.

To precisely estimate the integral representing each pixel, an oversampled numerical grid is required (see Figure \ref{fig:trpz} and paragraph \textit{Grid sampling} of Section \ref{sec:considerations}). Hence, for a given pixel, the solution of equation \ref{eq:solution} would be highly degenerate. In many situations, the system will still be invertible since many pixels can cover slightly different wavelength ranges, but the solutions will then be extremely unstable. This is an ill-conditioned system, where a slight change in the observation vector $\mathbf{D}$ could cause large differences in the solution $\mathbf{f}$. To circumvent this problem, the system needs to be regularized.

\subsection{Regularization} \label{sec:regularization}

There are multiple ways to perform regularization. The one chosen here is Tikhonov regularization \citep{tikhonov.1963}. It is also referred as Phillip-Twomey's regularization \citep{phillips.1962} and ridge-regression \citep{horel.1962}. This technique has been used in astrophysics in a variety of similar situations requiring the inversion of an integral equation \citep[e.g.,][]{kunasz.1973,thompson.1990}. In fact, it is part of some advanced spectral extraction methods for fiber-fed spectrographs that were identified among the most effective methods \citep{min.2020}. It is also used in the context of spline interpolation of noisy data \citep{green1993, hastie.1990}.  The main idea is to add a regularization term to the linear $\chi^2$ (equation \ref{eq:chi2}), yielding the following equation:
\begin{equation}\label{eq:tikhonnov}
    \chi^2_{\rm Reg} = \chi^2  + \alpha \left\| \Gamma \textbf{f}  \right\|^2 \, .
\end{equation}
Here, $\alpha$ is a Lagrange multiplier and $\Gamma$ is a linear operator that adds a ``cost" depending on the nature of the solution. Generally, it is used to favor smoother solutions, and hence reducing the level of overfitting. We can obtain the new solution by minimizing the system in the same fashion as before, differentiating with respect to $f_k$, with the following result:
\begin{equation}\label{eq:tikho_solution}
    \left(\frac{\textbf{B}}{\mathbf{\sigma}}\right)^T  \left(\frac{\textbf{D}}{\mathbf{\sigma}}\right)  
    = \left[ \left(\frac{\textbf{B}}{\mathbf{\sigma}}\right)^T
      \left(\frac{\textbf{B}}{\mathbf{\sigma}}\right)
      - \alpha \Gamma \right]
      \mathbf{f}_{\rm N_k} \, .
\end{equation}
In this case, the Tikhonov matrix $\Gamma$ was chosen to be the first derivative operator as in \cite{lamost_tikhonov.2015} or \cite{piskunov_optimal_2021}.
The choice of $\alpha$ can be done in many different ways. Usually, the general idea is to find a good balance between the regularization term and the $\chi^2$ which can be done using the L-curve criterion (e.g., \citealp{hansen1992}) or generalized cross-validation techniques, GCV \citep{golub.1979, wahba.1977}. The L-curve technique has the advantage of being robust, especially against correlated noise \citep{hansen1993} compared to the GCV. However, it tends to over-smooth the solution, which is not optimal. As for the GCV, it is too computationally intensive in the context of large-scale problems. However, in the present situation, the objective is not the direct product of the regularization, i.e., the underlying flux, but rather the modeling of the pixel after re-integration (this will be discussed in sections \ref{sec:considerations} paragraph \textit{Proper output} and \ref{sec:implementation}). Thus, it is not necessary to find the most physically accurate solution. In fact, what is needed is the smoothest solution that, once re-projected on the detector, fits the observations within the uncertainties. Consequently, we defined custom criteria to determine $\alpha$ with the objective to keep the dependency of the solution's sensitivity to the scaling factor lower than the expected noise. Mathematically, this comes back to defining a threshold on the derivative of the $\chi^2$ with respect to $\log \alpha$. The same convergence criterion was used by \cite{khmil.2002} in a similar context.

\subsection{Other considerations} \label{sec:considerations}

\begin{figure}
\centering
\includegraphics[width=\linewidth]{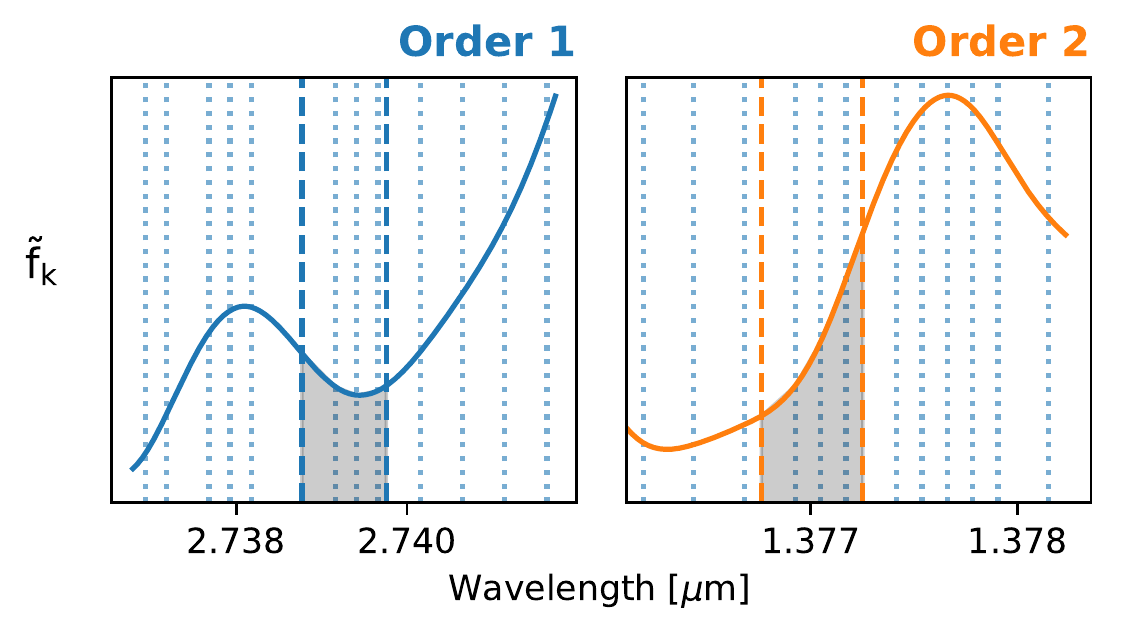}
\caption{Example of the trapezoidal integration on a specific grid. The shaded grey region represents the area under the curve for a trapezoidal integration over the wavelength coverage of a given pixel delimited by the vertical dashed lines. The vertical dotted lines indicate the irregular grid used to do the integration. The flux at each order's resolution is given by the solid lines.
}
\label{fig:trpz}
\end{figure}

\paragraph{Grid sampling} One important aspect of the technique is the choice of the wavelength grid. Since the simulated pixels are the results of numerical integrations, they are subject to computational errors. One way to define the grid would be to use a grid representative of the native pixel sampling and to oversample each interval of this grid by a certain factor. This will however create unnecessarily large systems and increase the computation time. Moreover, for the regularization method to be well-behaved, it is better to have errors of the same order for each node. Indeed, as highlighted by \cite{puetter_digital_2005}, the dynamical range needs to be constrained, lest some regions will be overfitted and others underfitted. Given all this, we opted for an irregular grid designed to make the magnitude of the integration error between subsequent nodes more uniform. To estimate the integration error on each node, we compared the result of a trapezoidal integration (as it is done in ATOCA) with a more precise Romberg's integration. The intervals with an estimated error higher than a specified tolerance were oversampled by a factor of two iteratively until the tolerance was satisfied. This method only requires an estimate of the function to be integrated, which can be given by a user or directly estimated from the data. An example of an uneven grid is shown in Figure \ref{fig:trpz}, represented by the vertical dotted lines. 

\paragraph{Proper output} As mentioned before, the underlying flux $\mathbf{f}$ is not the end product of an extraction. Indeed, since it has a resolution higher than both orders on the detector, $\mathbf{f}$ will be degenerate. In fact, solving for $\mathbf{f}$ is a deconvolution, which is subject to instabilities or artifacts \citep{bolton.2010}. Thus, one will have to reconvolve the result to get rid of these effects. In this case, the underlying flux has to be integrated on bins representing a pixel grid. This can be done by invoking equation \ref{eq:B} and reconstructing the detector, which can be used to assess the quality of the detector modeling by examining the residuals. It could also be used to rebuild each order independently using the $\textbf{b}_n$ matrices from equation \ref{eq:bn}. Furthermore, it is possible to get a one-dimensional spectrum by re-integrating on a grid representative of the pixel sampling. In fact, this is equivalent to reconstructing a single row of the detector. In this context, without any dimension in the cross-dispersion axis, the spatial profile would not be relevant anymore and its value should be set to unity.
%

\paragraph{Wavelength distortions} In a variety of situations, the wavelength solution will not be constant along the axis perpendicular to the dispersion. Generally, this can be due to differential refraction in the Earth's atmosphere and imperfect spectrograph optics \citep{horne.1986}. This effect is seen in many spectrographs \citep[e.g.][]{bolton.2010, piskunov_optimal_2021}. It can also be caused by observing techniques like the spatial-scan mode of the \textit{Hubble Space Telescope}'s Wide Field Camera 3 \citep{deming.2013}. In the case of the NIRISS SOSS mode, a tilt is present due to the grism configuration (Albert et al., in prep.).  Standard extraction procedures will either neglect this distortion \citep{sing.2015} or resample the detector image by interpolating along the dispersion axis \citep[e.g.,][]{kreidberg.2014}. ATOCA has the advantage of implicitly accounting for this distortion by treating each pixel individually, and using the full 2D wavelength solution.



\section{Implementation} \label{sec:implementation}

\begin{figure*}%
\centering
\includegraphics[scale=0.6, trim={3cm 2cm 3cm 2cm},clip]{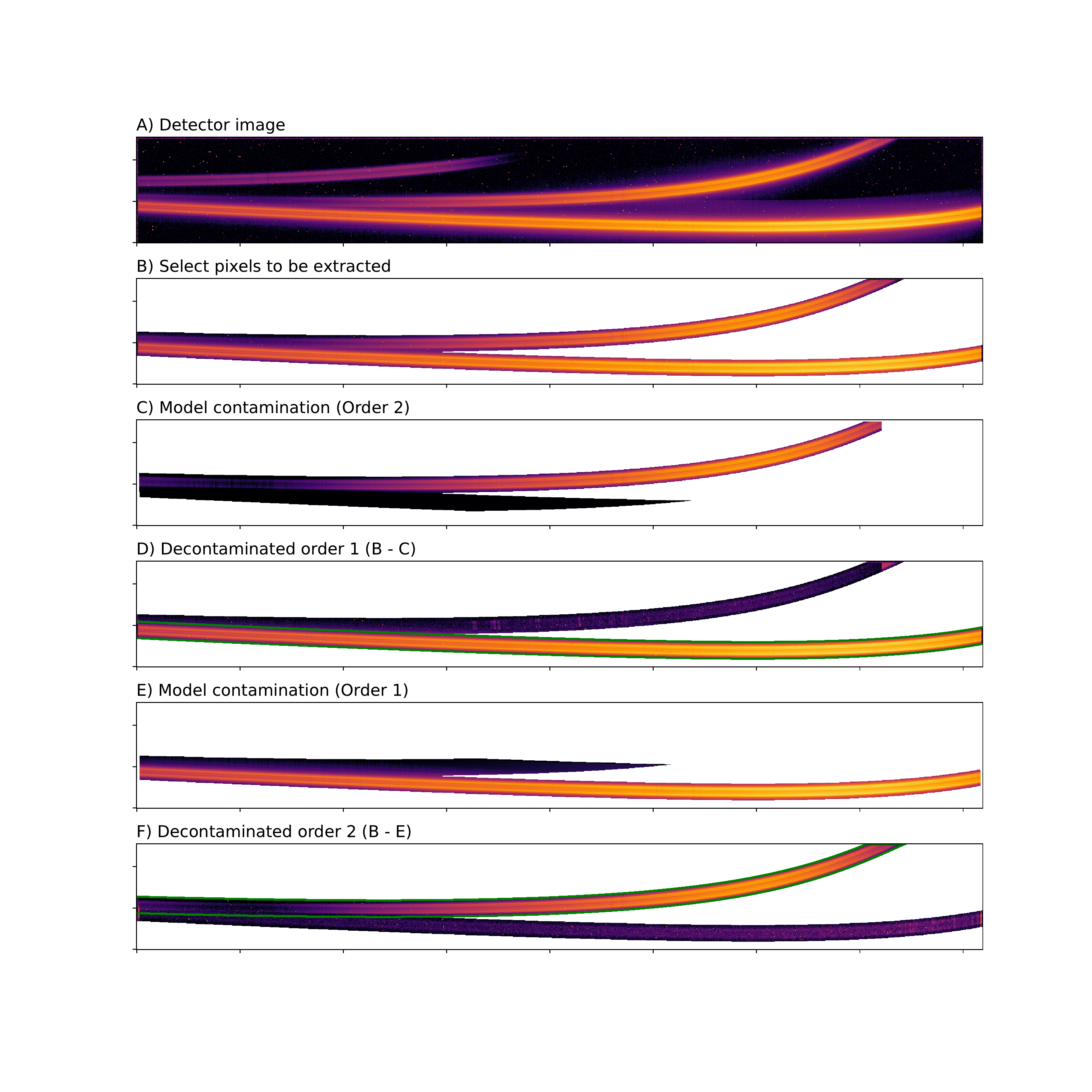}
\caption{\label{fig:steps}
Decontamination steps for the first and second spectral orders. From top to bottom: A) processed image of the detector with only signal from the targeted object, B) selection of relevant pixels, C) reconstructed image of the contaminating orders after a fit using ATOCA, D) decontaminated image of the first order after subtraction of the contaminating order; the data is now ready for a standard box extraction around first order (delimited by the green lines). Panels E and F are the same steps as C and D, but the extraction of the second order. The color scale is logarithmic, with the lowest intensity in black and the highest in yellow. The pixels that are not considered in the analysis are in white.}

\end{figure*}

ATOCA has been implemented in \texttt{python 3.8} as an option for the spectral extraction of NIRISS/SOSS observations in the JWST Data Management System (DMS). It is part of the \texttt{Extract1dStep}\footnote{\url{ https://jwst-pipeline.readthedocs.io/en/latest/jwst/extract_1d/arguments.html}} of the \texttt{spec2pipeline} calibration step. A description of the inputs and an example Jupyter Notebook can be found at \url{https://github.com/AntoineDarveau/atoca_demo}.

However, the end product is not the spectrum extracted with this method, as described in the Section \ref{sec:considerations} paragraph \textit{Proper output}. Indeed, one caveat regarding the product of an extraction with ATOCA is the dependency on the accuracy of the model. Just like the optimal extraction, the method needs a relative precision on the spatial profile much better than the corresponding data in order to make an unbiased spectral extraction \citep{horne.1986}. Moreover, inconsistencies between the spectral orders in the wavelength solution, the throughput or the resolution kernels could result in an intermediate solution between the two orders that will not satisfy the expected accuracy. However, in the case of contaminated spectra, it is possible to circumvent this problem by reconstructing the trace of each individual order with the $b_n$ matrices of equation \ref{eq:bn}, which are then used to decontaminate the detector image; the process is shown in Figure \ref{fig:steps}. This allows for a more classical extraction, like a box extraction, to be performed afterwards on each decontaminated trace. This technique has the advantage of reducing the dependency on the accuracy of the model. For instance, as shown in Figure \ref{fig:steps}, the modelling will only affect contaminated columns of the first order, leaving the rest untouched. This technique was preferred in the context of NIRISS/SOSS mode, although the ATOCA spectra are also available as a byproduct of the detector fit. Another output of the ATOCA is the model of each order (1st and 2nd) that were used for the decontamination. This could be useful to assess the quality of the fit by simply looking at the residuals. It would also be required to quantify the level of contamination (see Section \ref{sec:the_problem}). Furthermore, this model of the detector can be used to assign values to bad pixels located inside the aperture without the need for a separate outlier correction routine.

As shown in Figure \ref{fig:steps}, only the well-behaved pixels that will be extracted are considered in the application of ATOCA. This is in order to reduce the computational time and to make sure that the detector fit is not biased by superfluous regions of the detector. However, even with pixel selection, the size of the matrix $B$ in equation \ref{eq:lin_sys} is still considerable, with a size of $N_{pixel} \times N_{k}$. For example, in the realistic scenario of a 40-pixel aperture and a reasonable oversampling of the wavelength grid (tolerance of $10^{-3}$ per pixel), the dimension of the matrix will be around $150,000 \times 6000 = 9\times 10^8$. To overcome this problem, we took advantage of the fact that most elements of the matrix are null. Indeed, each order (or source) modeled by the matrix $B$ is represented as a block-diagonal that can be shifted with respect to the main diagonal. This enabled us to use the \texttt{scipy} \citep{scipy.2020} sparse matrices and drastically lessen the computational time and the amount of memory needed.

The implementation required the addition of new reference files to the Calibration Reference Data System (CRDS)\footnote{\url{https://jwst-crds.stsci.edu/}} used by the DMS, since ATOCA needs, for each order, the two-dimensional wavelength solutions, the spectral resolution kernels, the spatial profiles, and the spectral throughputs. The one-dimensional wavelength solutions and the throughputs are already a product of the standard calibrations, so the two-dimensional wavelength maps can be created simply by applying the tilt to the existing solution. The convolution kernels were determined from monochromatic point spread functions generated with \texttt{WebbPSF}. The resulting images were rectified to remove the tilt and summed over the spatial axis to keep only the spectral dependency. The most demanding input is still the determination of the spatial profile, but techniques typically used for optimal extractions could be applied directly to most of the spectral ranges, except for the overlapping parts. An algorithm to estimate the spatial profiles of both orders within the contaminated region is currently in development (Radica et al., in prep), and will be made available to complement ATOCA before the release of Cycle 1 data. 

As mentioned in section \ref{sec:considerations}, ATOCA needs an estimate of the underlying flux $f_k$ to generate an oversampled grid. In the context of the NIRISS/SOSS mode, this is done by extracting the underlying flux $f_k$ for each order separately with only the most contaminated pixels masked. These rough extractions are done over a grid at native pixel sampling, which ensures the stability of the solution and removes the need for any regularization. Precision will be lost in this process and the contamination will not be treated correctly, but it is sufficient to generate the estimate. The latter is also used to provide a first estimate of the regularization factor, which is refined afterwards as described in section \ref{sec:regularization}. This step can take some time depending of the precision needed for the oversampled grid and the number of relevant pixels. Fortunately, it only needs to be performed once for a given time series, since the underlying spectrum will not vary enough to justify a different level of regularization.


In realistic observations, the position of the trace will change slightly between visits due to the angle of the pupil wheel and variation in the target acquisition, which will alter the wavelength solution as well as the spatial profile. These changes will effectively take the form of a rotation and a spatial shift of these reference files and are implemented as input parameters (rotation and translation) captured by the keyword \texttt{soss\_transform = [x, y, theta]}. These parameters can be either specified by the user or determined within \texttt{Extract1dStep} by fitting the measured trace centroid.


\section{Validation} \label{sec:validation}

\subsection{Simulations} \label{sec:simulation}

Two types of simulations were used in the context of this work: simulations produced with ATOCA itself (ATOCA simulations), and those produced by the instrument development team (IDT simulations). The first type takes advantage of the ATOCA capability to directly model the detector image using equation \ref{eq:lin_sys}, given an input spectrum. Hence it is possible to use it to generate simplistic simulations to validate the internal consistency of the method. To make sure that the numerical precision was not an issue, the wavelength grid used for the simulation was oversampled to limit the integration error over each pixel below $\sim$1\,ppm. We only included photon and background noise, so neither bad pixels, 1/f noise, nor cosmic ray hits were taken into account. The reference files were based on the best current knowledge of the NIRISS/SOSS mode, as described in section \ref{sec:implementation}. We used the throughput and the one-dimensional wavelength solution estimated or measured in the lab by the NIRISS instrument development team. The spatial profiles were determined from the IDT simulations

The IDT simulations (see Albert et al, in prep) are made by distributing an incident flux directly on an oversampled image using a trace as wide as a single oversampled pixel. This signal is then convolved in two dimensions with a grid of monochromatic kernels from WebbPSF and re-sampled at the native pixel resolution. These simulations were used to test the versatility and robustness of ATOCA on more realistic simulations. They also include the 1/f noise, the effects of flat-fielding and bad pixels. In all situations, the incident stellar fluxes are taken from high-resolution PHOENIX synthetic spectra \citep{husser.2013} and the transit models from SCARLET \citep{benneke.2015}.


\subsection{Results} \label{sec:results}
\begin{figure*}
\centering
    \begin{subfigure}
    \centering
    \includegraphics[scale=0.6,trim={0.5cm 2cm 5.7cm 2cm},clip]{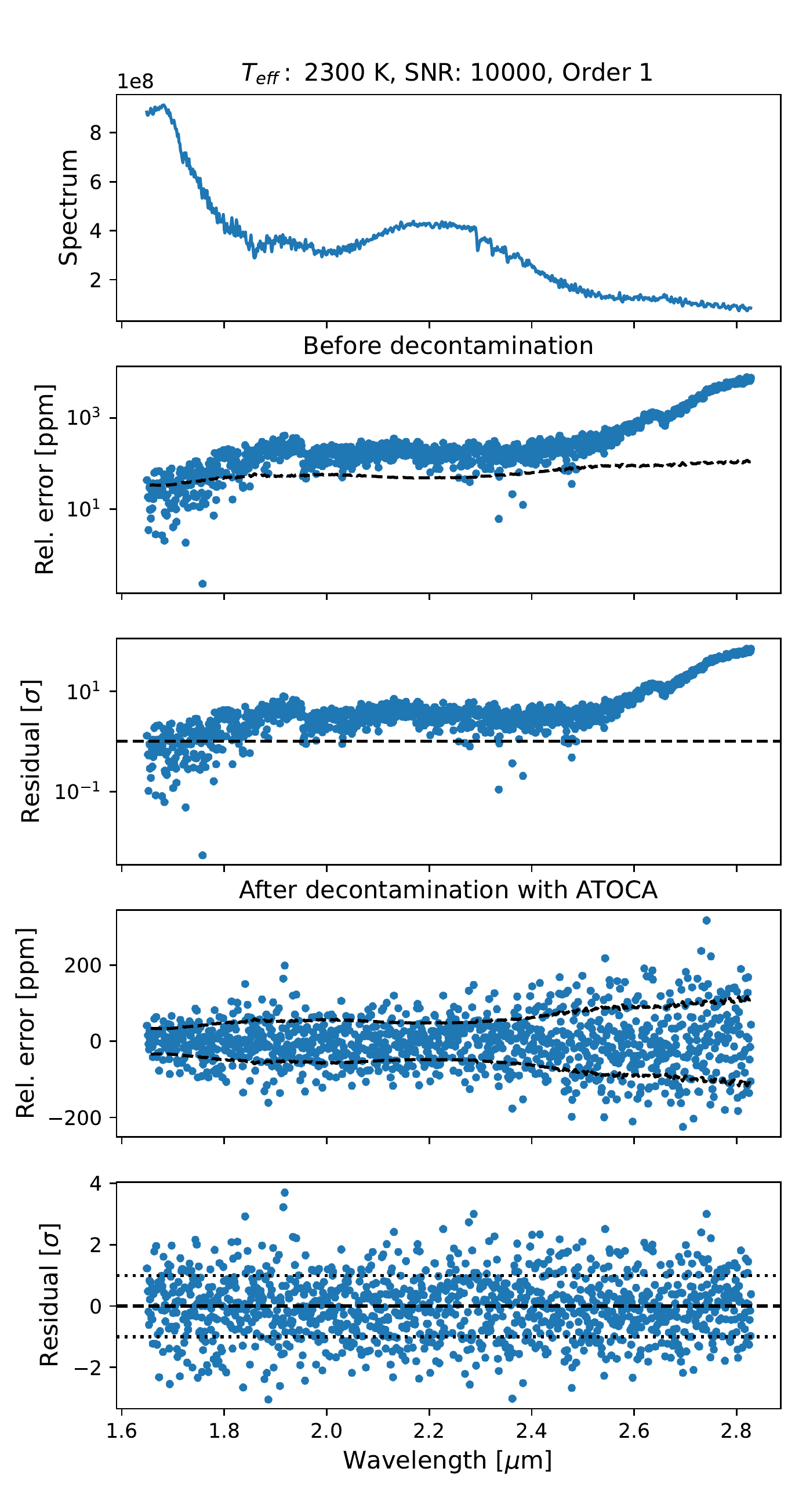}
    \end{subfigure}
    \begin{subfigure}
    \centering
    \includegraphics[scale=0.6,trim={0.5cm 2cm 5.7cm 2cm},clip]{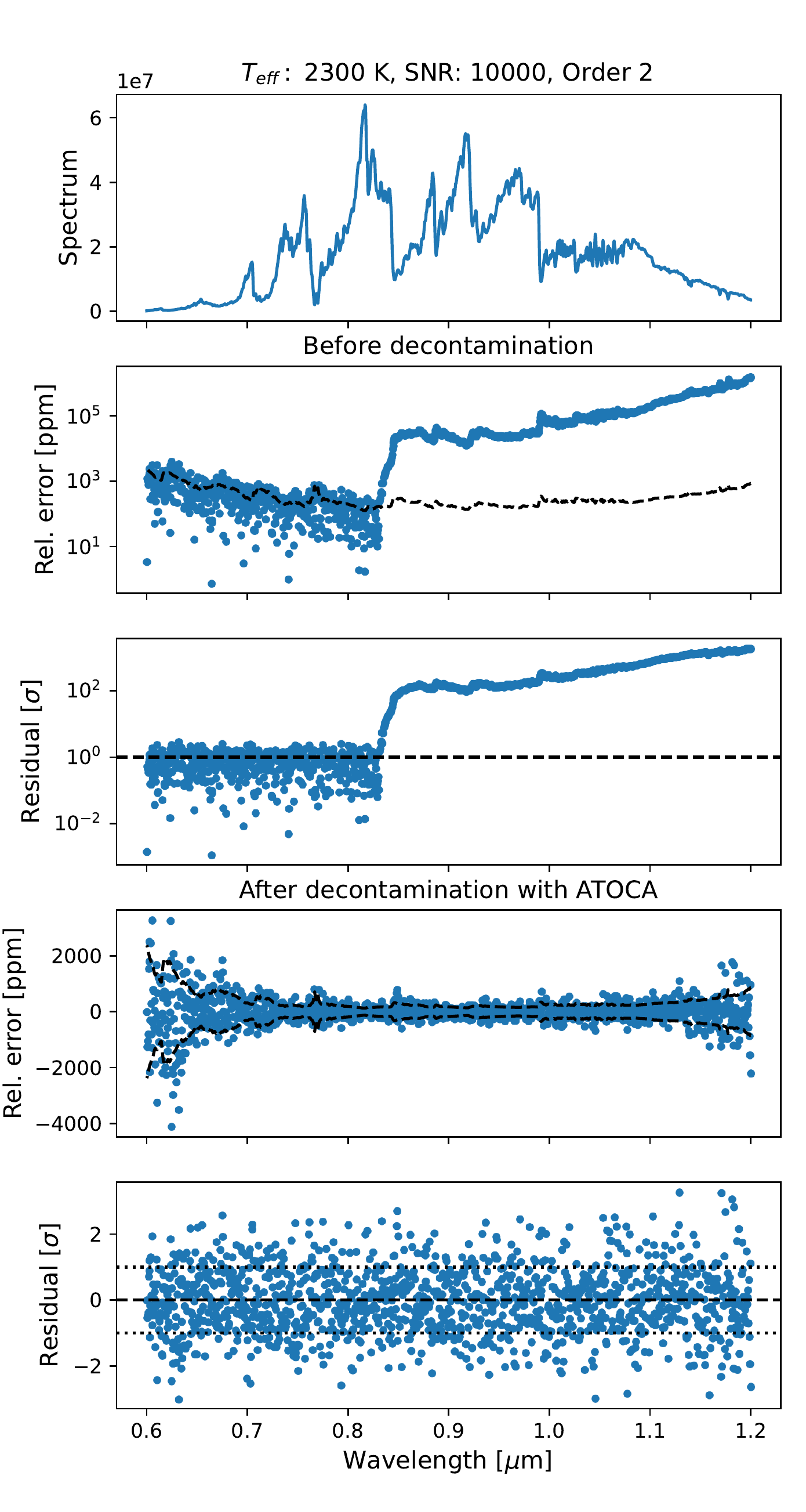} 
    \end{subfigure}
    \caption{Decontamination on a single image. The extracted spectra are shown in the top panel. The four other panels show the extraction residuals before and after application of ATOCA, at two different scales. The dashed curves correspond to the expected 1-$\sigma$ uncertainties (in absolute value for the panels 2 and 3).
    For the last panel, the 1-$\sigma$ thresholds are marked by a dotted line. The simulation is the equivalent of a deep stack with a signal-to-noise ratio of 10,000 for a 2300\,K star.}%
    \label{fig:consistency2300}
\end{figure*}


\begin{figure*}
\centering
    \begin{subfigure}
    \centering
    \includegraphics[scale=0.70]{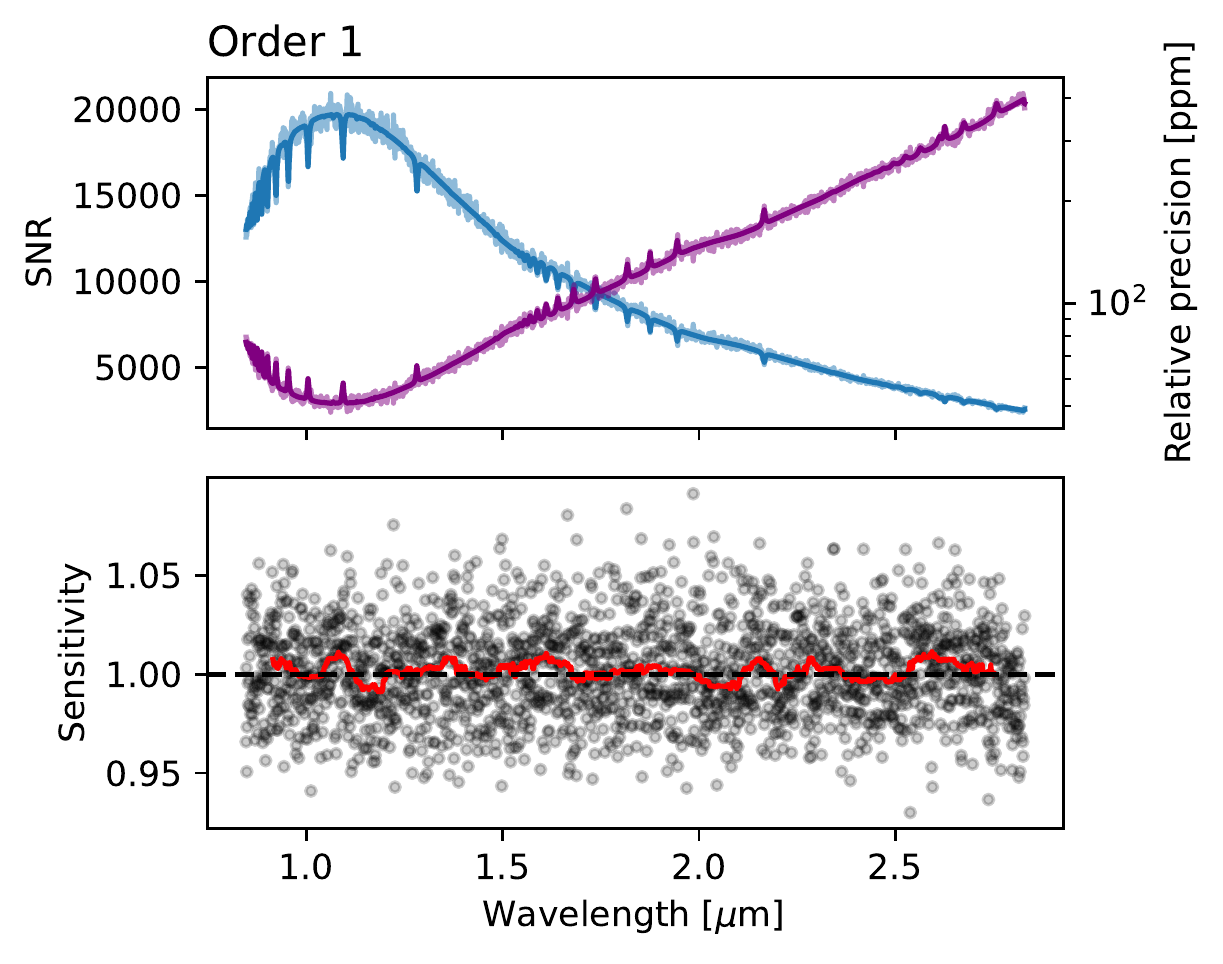}
    \end{subfigure}
    \begin{subfigure}
    \centering
    \includegraphics[scale=0.70]{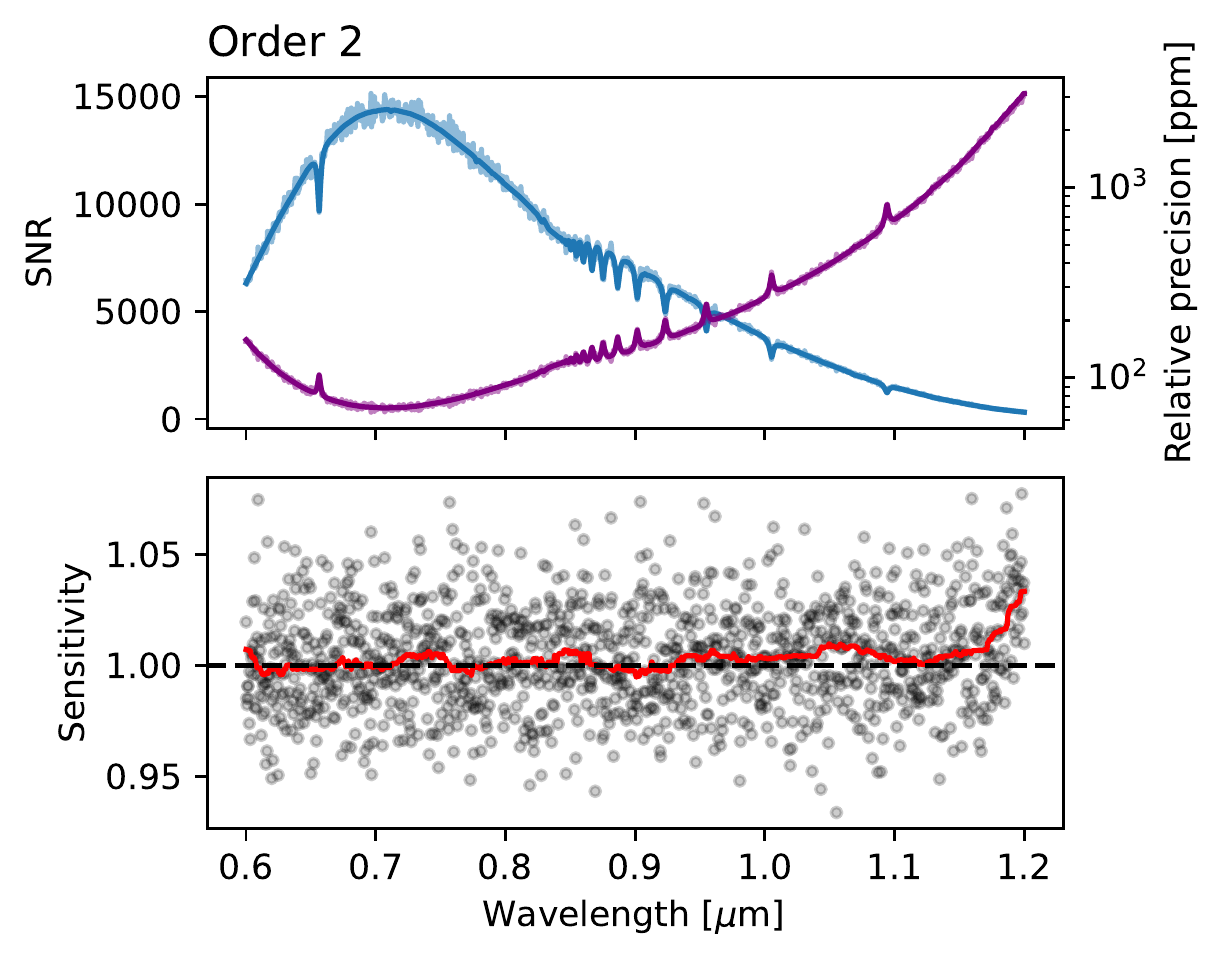}
    \end{subfigure}
\caption{Stability of the decontaminated extraction. The top panel displays the expected signal-to-noise ratio (opaque blue) and the corresponding precision (opaque purple) of the time-series observation of BD+601753. The standard deviation along the 876 box-extracted spectra is used to determine the measured signal-to-noise and precision (light blue and light purple). The sensitivity shown at the bottom panel is given by the ratio of the measured and expected scatter. A median filter of 81 pixels is represented by the red curve.}
\label{fig:stability9400}
\end{figure*}


\begin{figure*}
\centering
    \begin{subfigure}
    \centering
    \includegraphics[scale=0.8]{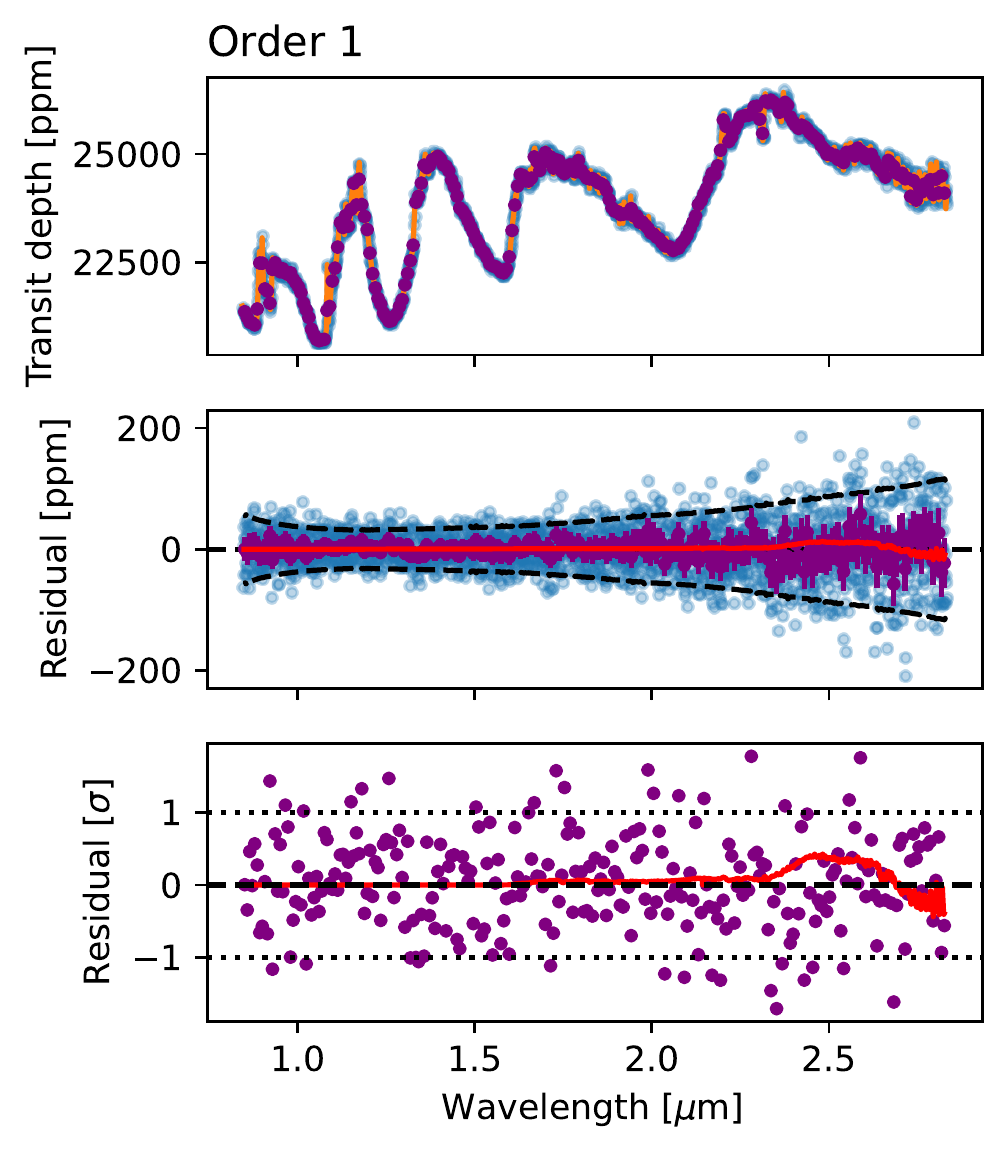}
    \end{subfigure}
    \begin{subfigure}
    \centering
    \includegraphics[scale=0.8]{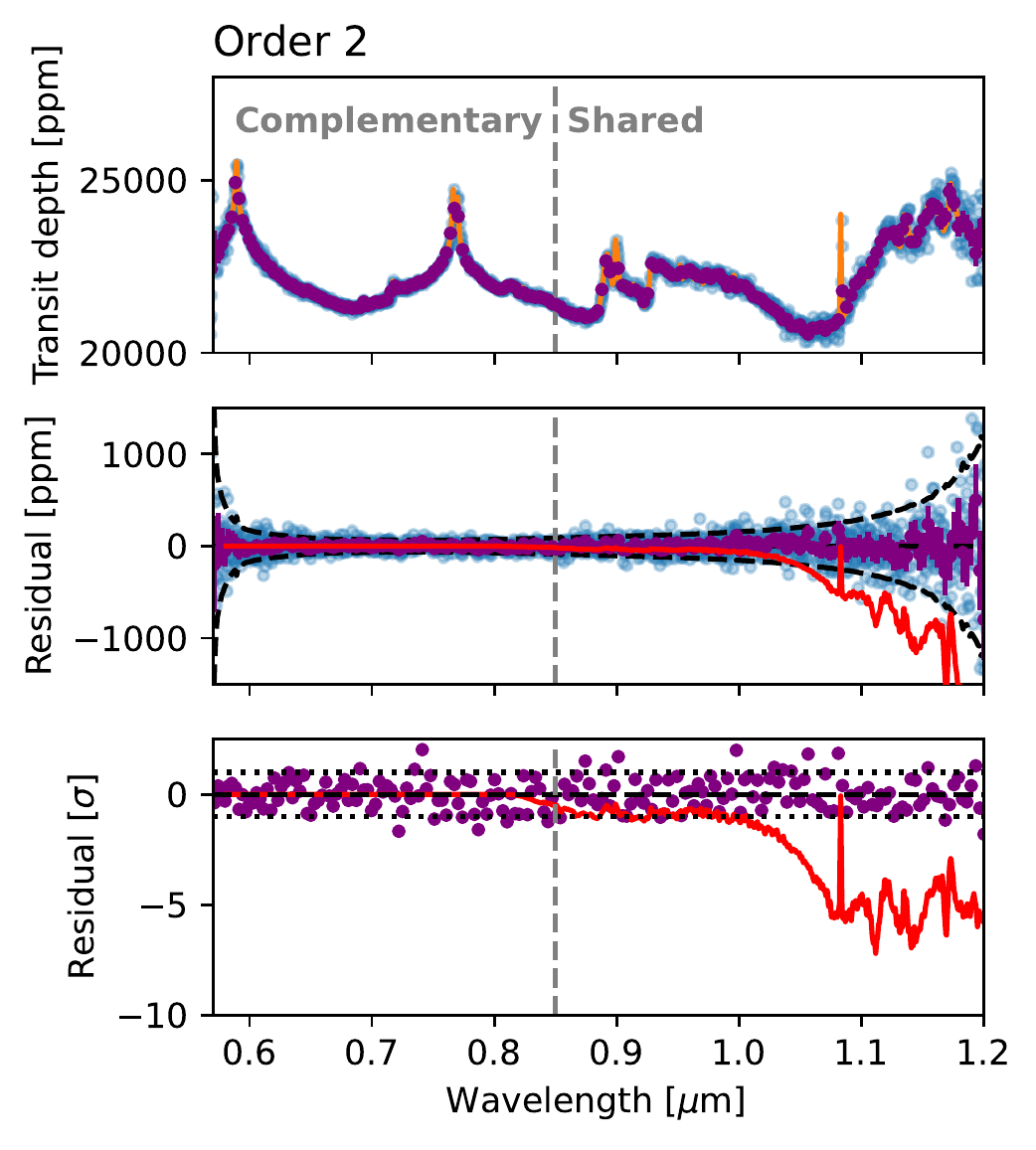}
    \end{subfigure}
\caption{Decontaminated transmission spectrum for a target similar to WASP-107\,b. The transmission spectrum as well as the residual in ppm and scaled to the uncertainties $\sigma$ are presented by the three panels. In each of them, the results are shown at native sampling in blue and with 8-pixel bins in purple. The dashed black lines indicate the expected value or the $1\sigma$ uncertainties. The estimation of the contamination from equation \ref{eq:transit_contamination} is plotted in red. In the first panel, the input transmission spectrum is also shown in orange. The vertical gray dashed line delimit the wavelength range from the second order that is complementary to (or shared with) the first order.}
\label{fig:hot_jup_tr}
\end{figure*}

\begin{figure*}
\centering
    \begin{subfigure}
    \centering
    \includegraphics[scale=0.8]{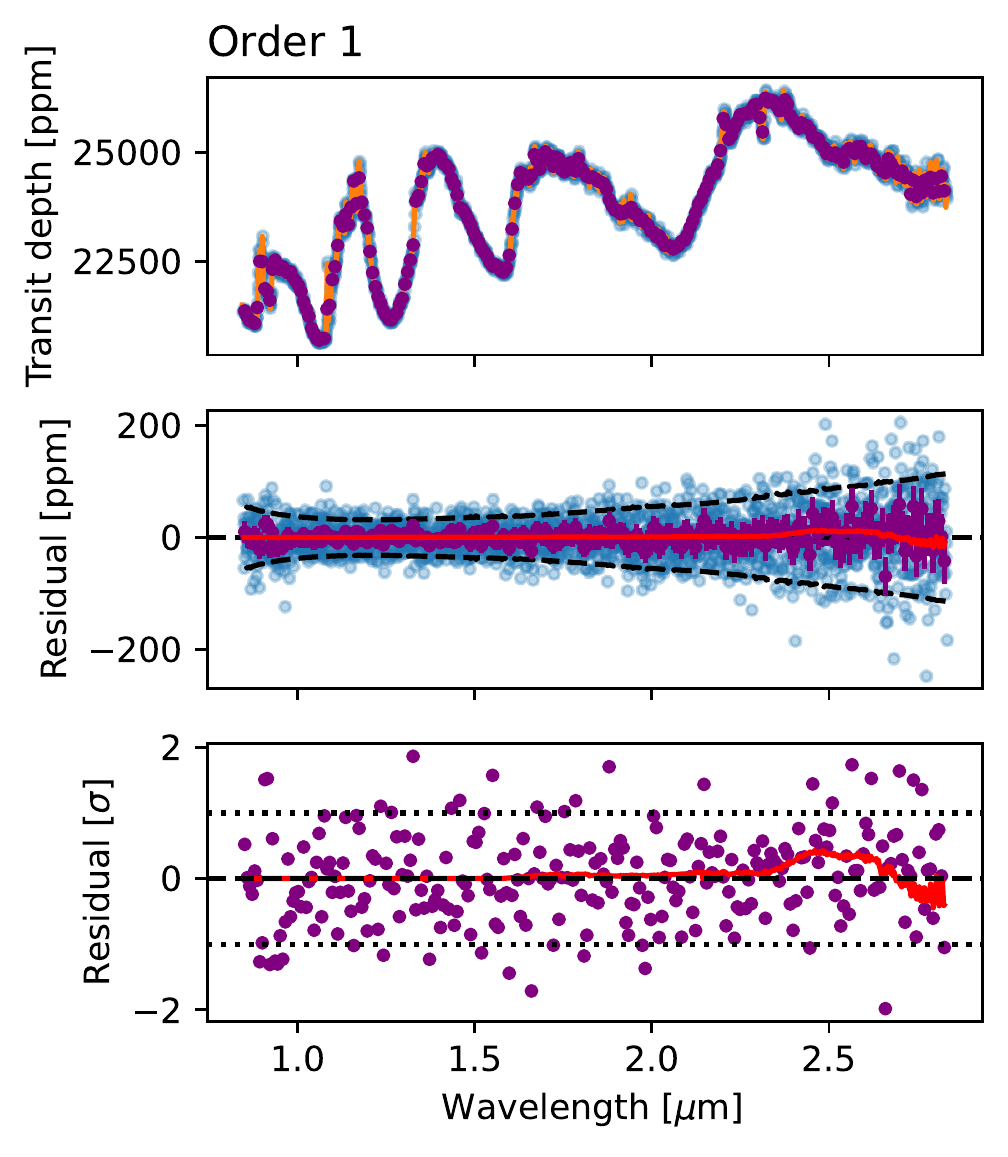}
    \end{subfigure}
    \begin{subfigure}
    \centering
    \includegraphics[scale=0.8]{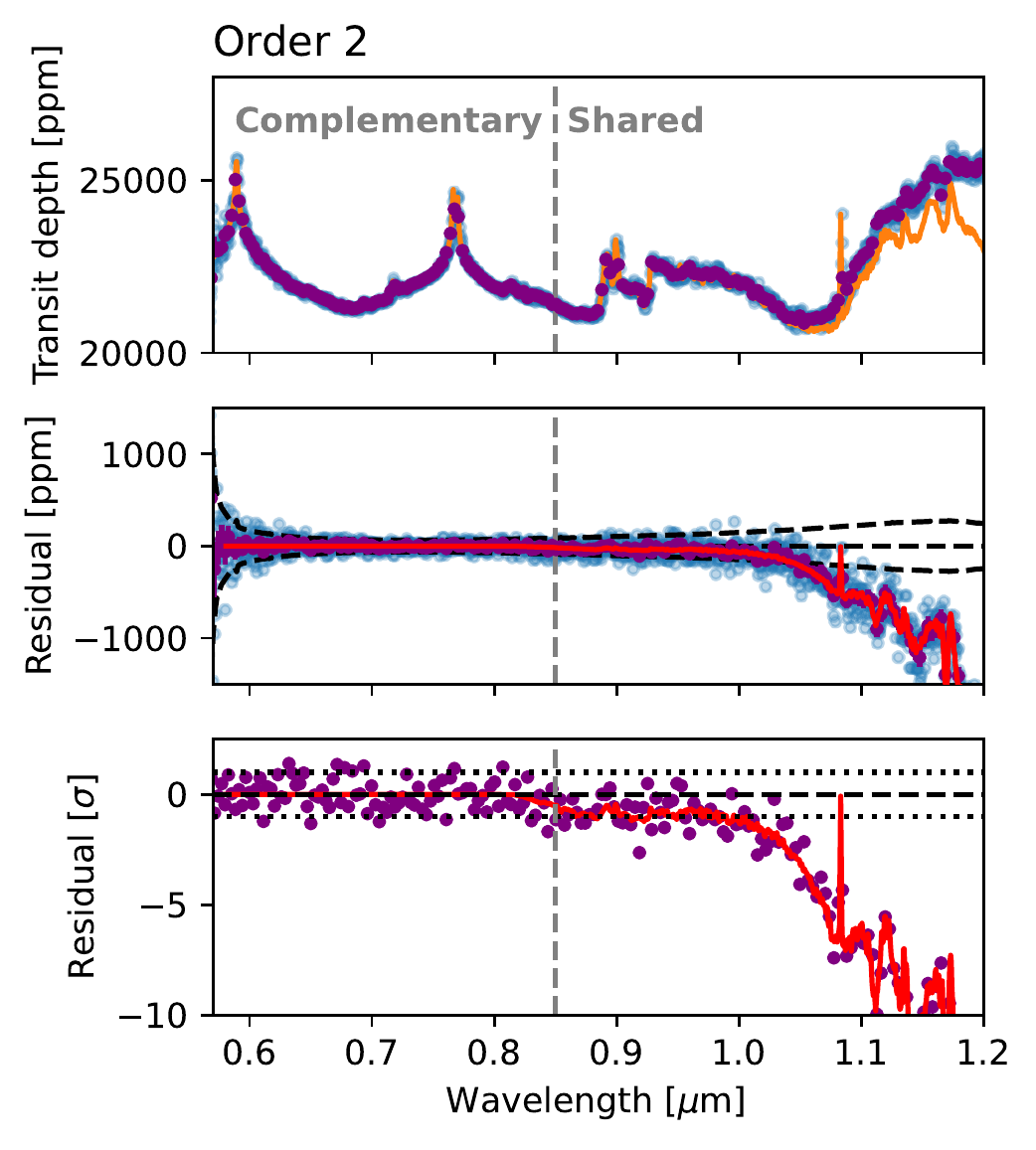}
    \end{subfigure}
\caption{Transmission spectrum for a target similar to WASP-107 b, without decontamination. Same description as in figure \ref{fig:hot_jup_tr}.  }
\label{fig:hot_jup_tr_cont}
\end{figure*}

\begin{figure*}[tbph]
    \centering
    \includegraphics[trim={0cm 0cm 2cm 0cm} , clip, width=\linewidth]{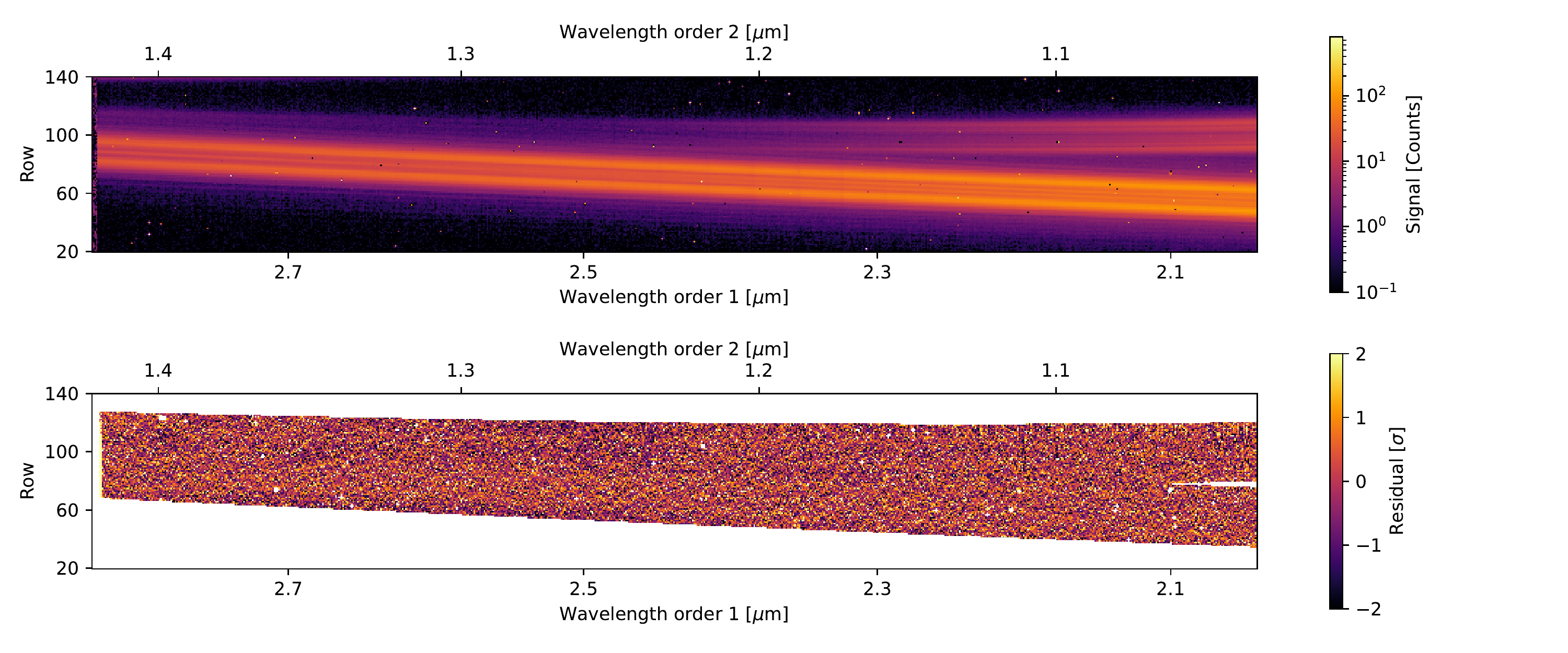}
    \caption{Comparison between the ATOCA modeling and the detector image of a single exposure. The top panel presents a close up of the region of the detector where the overlap occurs, corresponding to the lower left corner of Figure \ref{fig:sossmode}. The residual between the model extracted by ATOCA and the simulated detector image is shown in the bottom panel. The color scale is in units of $\sigma$, the pixels uncertainty.  The pixels that were not used for the fit (e.g., bad pixels or background) are in white. }
    \label{fig:residual}
\end{figure*}

\begin{figure*}
\centering
    \begin{subfigure}
    \centering
    \includegraphics[scale=0.68]{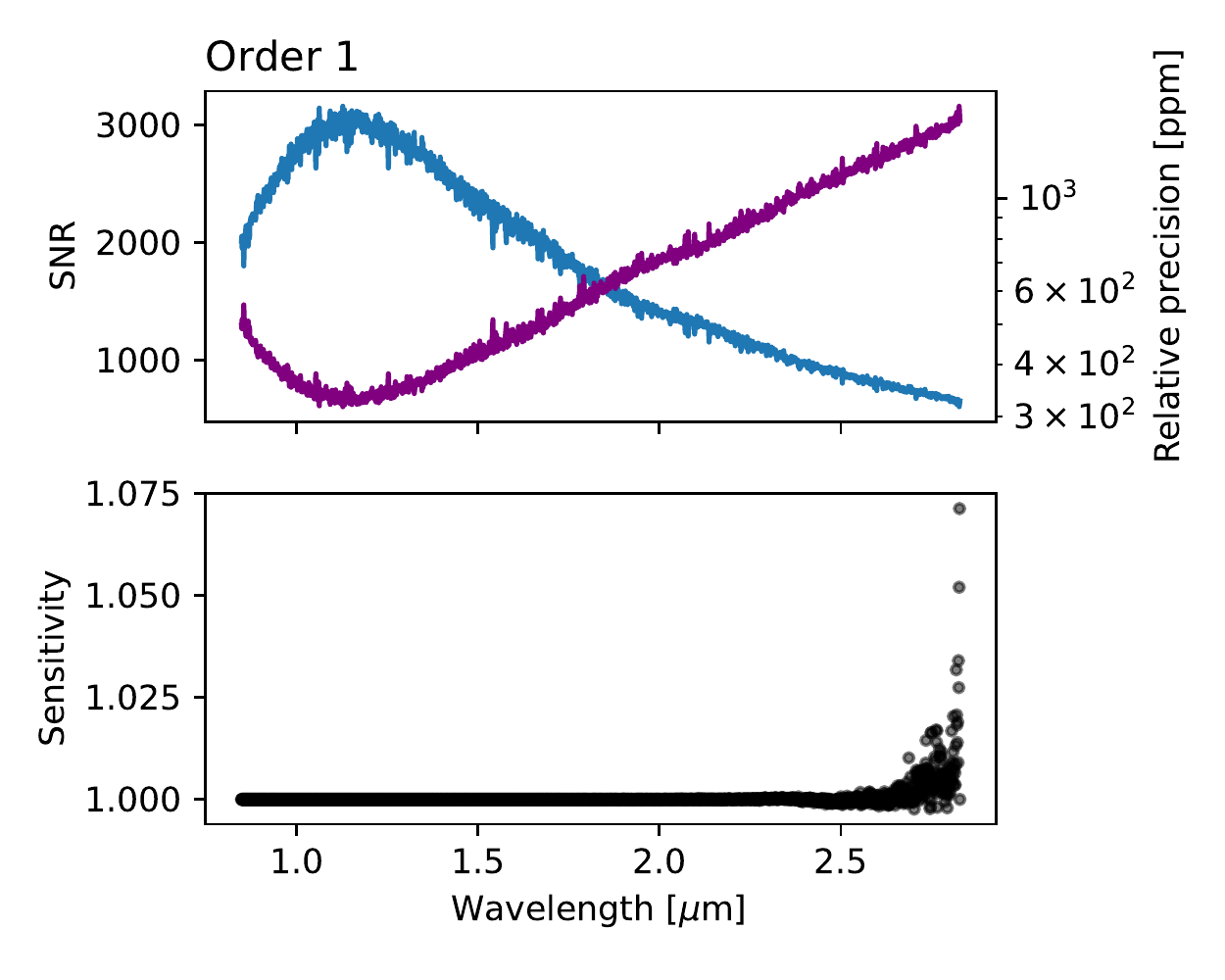}
    \end{subfigure}
    \begin{subfigure}
    \centering
    \includegraphics[scale=0.68]{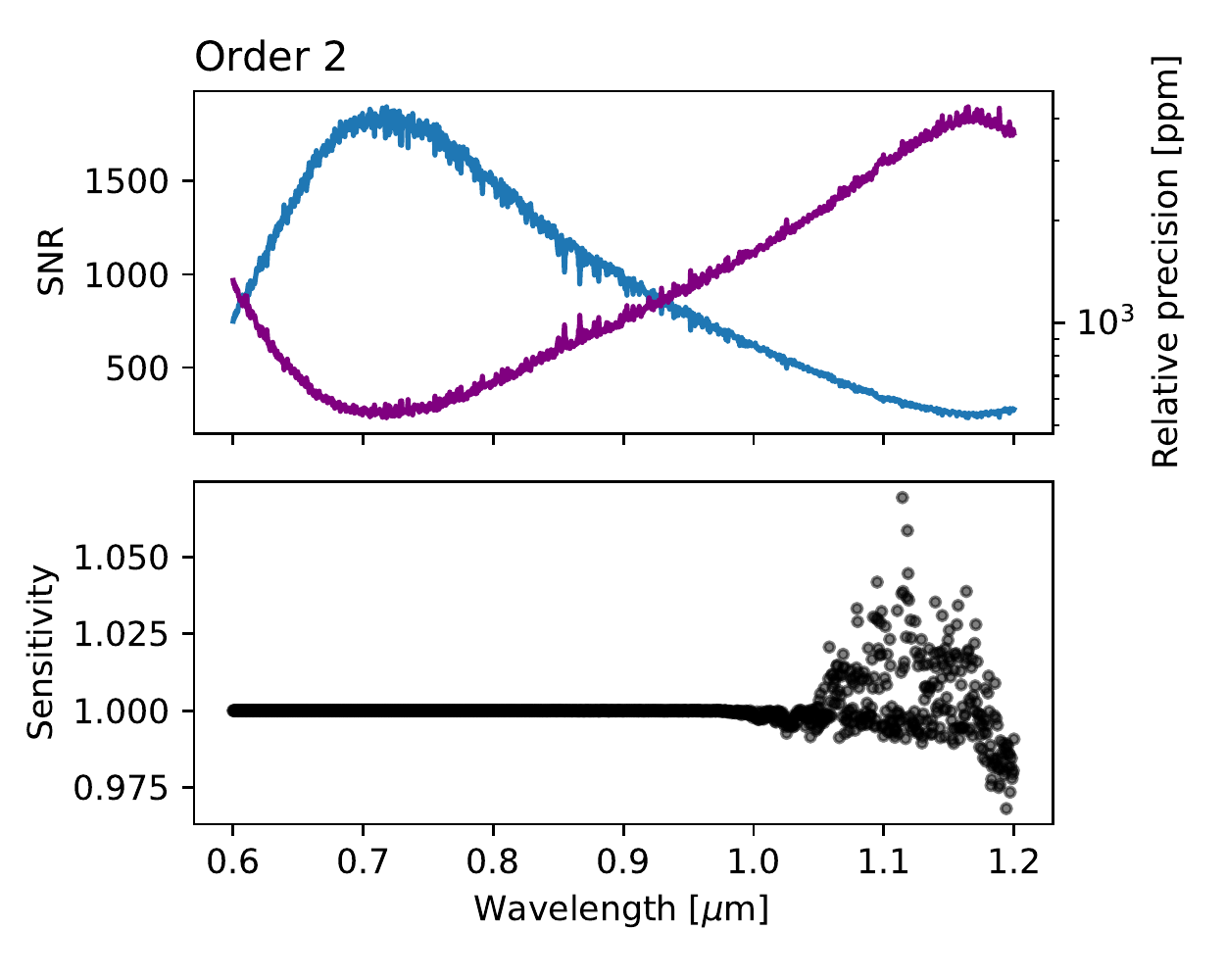}
    \end{subfigure}
\caption{Stability of the decontaminated extraction. The top panel displays the expected signal-to-noise ratio (blue) and the corresponding precision (purple) of the time-series observation of WASP-52 b. The sensitivity shown at the bottom panel is given by the ratio of the measured scatter before and after decontamination.}
\label{fig:stabilityWASP52}
\end{figure*}

For the first series of validations, the simulations were made with ATOCA itself to assess the performances of the decontamination and internal consistency. For each decontamination, the estimated relative tolerance of the wavelength grid had to be less than $10^{-3}$, a level of precision that implies a reasonable grid length, and hence moderate computational time. We also do not expect the precision of a single pixel to get higher than this, which corresponds to a SNR of 1000.

A first test was done on a single exposure by comparing each extraction to an equivalent simulation free of contamination. Different maximum pixel signal-to-noise ratios were tested, ranging from 200 to 10,000. This verifies our ability to decontaminate observations with the most extreme levels of precision planned for the NIRISS/SOSS mode. An example of the results is shown in Figure \ref{fig:consistency2300} for a star with an effective temperature of 2300\,K and for a SNR of 10,000. The decontamination performance proved to be very effective at removing the contamination from the second order. In the case presented in Figure \ref{fig:consistency2300}, ATOCA was able to go from a contamination of $\sim10000$\,ppm, equivalent to a hundred times the expected uncertainties, to virtually no contamination. The residuals fall within the expected uncertainties ($<$100 ppm) and seem free of any correlated noise. The same conclusions hold for all stellar temperatures. It is also interesting to note the clear cut around 0.85\,\micron\ in the contamination levels (before decontamination). This is an artefact of the simulations since the monochromatic kernels used for the two-dimensional convolution only cover 128 native pixels. Thus, there is a threshold at 69 rows around the center of the trace where the wings of the spatial profile are not modeled. In the context of real observations, the contamination levels should extend below 0.85\,\micron, while continuing to decrease.

The second and third tests were inspired by the commissioning programs \cite{1091jwst} and \cite{1541jwst}. The former is a flat time series observation comprising 876 integrations on the standard A1V star, BD+601753. This will quantify the stability of frame-by-frame decontamination for representative SNRs (177 at maximum pixel). The results are presented in Figure \ref{fig:stability9400}. The flat sensitivity spectrum shows that the frame-by-frame decontamination is stable. The combined spectrum reaches a precision of less than 100 ppm at best and between 150 and 400 ppm in the regions subject to contamination. The measured standard deviation along the time series is in good agreement with the expected uncertainties in the entire spectral range, as evidenced by the sensitivity curve. Note that these results only accounts for photon noise; in realistic observations, other sources of noise like 1/f, jitter or other detector effects might become dominant at certain wavelengths.

The latter program consists of another time-series observation of an expected featureless transit, evaluating the precision of relative measurements. Our simulation was based on the primary target of this program, the massive hot-Jupiter HAT-P-14\,b, which was simplified to a step-transit (no limb darkening, instantaneous ingress and egress) with a flat transmission spectrum. We also neglected the effect of non-linearity in the ramps and forced a signal-to-noise ratio of 400 per pixel at maximum. This is above the capability of a SOSS-mode single integration, but it allows to push the decontamination at higher levels. The same framework was also applied to a transit spectrum of an exoplanet similar to WASP-107\,b in a fourth verification (same transmission spectrum and same star, but different magnitude). In this case however, the signal-to-noise was artificially increased to 1000 for the same reason as mentioned above. Both validations led to the same conclusions, so the results of the flat transit are not shown here. The WASP-107\,b-like transmission spectrum is presented in Figure \ref{fig:hot_jup_tr}. It confirms that the procedure can reach the expected precision and accuracy on relative measurements. Even with a required precision on ATOCA's wavelength grid of $10^{-3}$ per pixel for the integration-by-integration decontamination, the combination of all extractions reaches an accuracy of less than 100 ppm in the order 1 contaminated region with, again, no evidence of systematic bias. The performance is even more obvious when it comes to the second order, where the contamination reaches levels of 1000 ppm. For comparison, Figure \ref{fig:hot_jup_tr_cont} presents the results of the WASP-107\,b time series without any decontamination. We can see that the polluting signal follows the expected curve (red) taken from equation \ref{eq:transit_contamination}. It also confirms that the levels of contamination for the first order are small compared to the actual signal. The result of the flat transit is not shown here since it leads to the same conclusions.

Based on these four tests, some remarks can be made. First, it is interesting to note that they entail qualitatively different high-resolution PHOENIX synthetic spectra \citep{husser.2013} at $T_{eff} = 9400 \rm \, K $ (BD+601753), $T_{eff} = 6700 \rm \, K $ (HAT-P-14), $T_{eff}=4500 \rm \, K$ (WASP-107) and $T_{eff} = 2300 \rm \, K $; the hottest spectrum showing well-defined absorption features on a smooth continuum and the coldest one containing features with noise-like behaviour. This is an assessment of the robustness of ATOCA regarding the nature of the underlying spectrum $f_k$.

Second, the tests were initially run assuming that the spatial profiles, the wavelength solutions, and the throughputs were exactly known, resulting in errors consistent with the expected noise limits for each application of ATOCA. However, the same tests were repeated with the reference files slightly shifted from their nominal values to confirm the robustness of the decontamination by applying a rotation and spatial shift (see end of Section \ref{sec:implementation}). We found no evidence that it affected the extraction for reasonable values, i.e., within the expected precision of the reference files. More precisely, we tested for shift in the dispersion direction up to 0.5\,pixel, for shift in the spatial axis up to 0.1\,pixel and for rotations up to $0.01\deg$.

Third, based on the apparent agreement between the expected and measured transit contamination seen in Figure \ref{fig:hot_jup_tr_cont}, it would seem that the correction for contamination could be made after a standard extraction, directly on the one-dimensional spectra using equation \ref{eq:transit_contamination}. However, these examples used an idealized transit, with a perfectly stable stellar spectrum and without considering any limb darkening. Further analysis should be done before making any conclusion on this possible alternative to correct for order contamination.

ATOCA was also applied on the realistic time series simulation from the IDT to test the robustness of the algorithm. We present here an example on WASP-52. In this case, the tests were designed to assess the quality of integration-by-integration decontamination as well as the stability of the decontamination. Only the stellar spectrum was included since adding a transit would only add complexity to the interpretation of the results, without bringing additional information. The time series comprises a total of 103 integrations with a signal-to-noise ratio per pixel reaching up to $\sim 250$. 
Contrary to the more simplistic simulations, we did not have access to each individual order, which is more representative of the context of real observations. Therefore, we used the residual of the full detector model (combined orders) for individual integrations as an indicator of the quality of the decontamination. The logic being that, if the model is able to represent correctly the overlapping region as well as the pixels covering the wavelength domain shared between both orders, then we can be confident that the overall model is accurate. Figure \ref{fig:residual} presents the residuals for a single integration, given by the equation

\begin{equation}
    \rm residual = \frac{observation - model}{uncertainty} \, .
\end{equation}

In this situation, since the input spectrum is perfectly stable, the uncertainties could be determined empirically using the standard deviation of each pixel throughout the full time series. The result is consistent with Gaussian noise and there is no evidence of any correlated features.

Concerning the stability of the decontamination, we had to do a similar comparison as in Figure \ref{fig:stability9400}. However, this time, the spectrum extracted with the standard technique could not be used directly to avoid a possible contribution from the bad pixel modeling. Instead, we used the standard deviation of the pixels along all integrations, as it was done above to estimate the uncertainties, and then computed the summation in quadrature with the weight specific to the extraction method. In this manner, the bad pixels are not included in the summation. This was done for the time series before and after decontamination. The results are presented in Figure \ref{fig:stabilityWASP52}. The sensitivity increases slightly at longer wavelengths where the contamination is at its peak, but it remains contained below 5\% for practically all of the domain of both spectral orders. This means that the precision, which is around 1000\,ppm in the current example as shown in the top panel, would differ by only 50\,ppm. 

Based on these two tests, we can conclude that ATOCA can model the detector image within the uncertainties and at a low cost in terms of noise. This shows again that the algorithm is robust to inexact reference files. 

On a different note, it is important to mention that the quality of the modeling is strongly influenced by the intertwining between both orders in their shared wavelength domain. This effect is accentuated in regions where the signal from both orders is strong, in which case a poor representation of the relation between the two can lead to over- and under-estimation. It can also be compensated with lower regularization factors, i.e., over-fitting. This was seen in many situations where the reference files were biased on purpose, as well as in the realistic simulations. This effect can be overcome by a proper estimation of the reference files. Thankfully, the regions that require higher precision are practically free of contamination, so the calibrations of the spatial profiles, the wavelength solutions and the throughput are relatively straightforward. Conversely, in the overlapping region, the throughput from the second order drops considerably, leaving the solution of the underlying flux dominated by the corresponding wavelength from the first order. This means that the model of the order 2 in the overlapping region is more permissive.

\section{Future improvements}

\paragraph{Hyper-parameters} The choice of regularization parameters could benefit from further improvements. Indeed, the current criterion could lead to unstable solutions, which would be very effective at modeling each valid pixel of the detector, but mediocre when it comes to accurately estimating any pixels that are not included in the fit. The latter objective could be achieved using other criteria. For example, non-exhaustive cross-validation techniques (e.g., k-fold, Monte-Carlo) would be a judicious choice since their primary objective is to be able to simulate a set of data points that are voluntarily excluded from the fit.

\paragraph{Choice of Tikhonov matrix} The injection-recovery tests on our simulations pointed towards comparable performances with the first and second derivative operator. The former was preferred due to its slightly lower complexity. However, the latter could end up being a more appropriate choice. Indeed, the second derivative is used in spline interpolation on noisy data to smooth out the solutions, which is not far from the problem we are facing here. It would also be a more physical explanation, given that the finite resolution of observations enforces a smooth solution with only small variations of the second derivative.

\paragraph{Background fitting} One forthcoming challenge with the NIRISS/SOSS mode is the background subtraction. While the spatial spread of the trace profile is very effective to improve the precision of the measurements, it also greatly reduces the number of pixels available to measure and remove the background contribution. This problem is even more concerning in the SUBSTRIP96 observing mode, where the two traces cover the entire range of rows for some columns. ATOCA could circumvent this problem by directly including the background in the fitting, adding the parameters needed to model the background at each column to the solution vector.

\section{Conclusion}

In this work, we presented an alternative spectral extraction method to solve the overlap problem pertaining to the NIRISS/SOSS mode. We first characterized the extent of the contamination for the first and second diffraction orders. It was found that for absolute measurements, the levels were kept below 0.1\% for most of the wavelength domain of the first order, except for wavelengths greater than 2.6~\micron\ where they reach 1\%. For the second order, the effect is much more important, but concerns mainly the wavelength range already covered by order 1. For relative measurements, for which the SOSS mode is specifically designed, the same levels of contamination are expected, but only on the chromatic differences of the signal (e.g., the difference in transit depth). This means that the systematic error due to the overlap should not be the dominant source of noise in the first diffraction order, although one should always assess its importance.

Nevertheless, it is still important to provide a way to disentangle each order's contribution to at least quantify the contamination, but also to allow a proper extraction for any absolute measurements or scenarios where the relative contamination becomes non-negligible. Consequently, we developed ATOCA, an algorithm that enables the modeling and extraction of overlapping orders (or sources), and decontamination the detector image. We showed that, given reasonable estimates of the spatial profiles, the wavelength solutions, and the spectral throughputs, ATOCA was able to decontaminate the data up to the required precision. We also characterized the robustness of the decontamination by introducing errors in the reference files and by applying it to realistic simulations. A first version of the algorithm is available in the JWST official pipeline. A development version is also available on github\footnote{\url{https://github.com/AntoineDarveau/jwst}}.

ATOCA is a promising technique to disentangle the contributions of overlapping spectral traces. Its framework might be transferable to other contexts, like field decontamination or multi-object slitless spectroscopy. ATOCA might also provide a powerful alternative to manage distorted wavelength solutions. All this potential has yet to be vetted throughout real observations which should come soon with the upcoming commissioning of NIRISS/SOSS.

\paragraph{Acknowledgments}
The name of the algorithm, ATOCA, is a word used in North-America French to designate the cranberry fruit. It was borrowed from the native American languages, possibly from Algonquin spoken by nations living in the area of the present-day Wisconsin \citep{canada} or from the Wandat word \textit{atokha} \citep{dhfq}. In this regard, we want to acknowledge the pivotal contribution of the First Nations to the North American French culture. We want to thank Anne Boucher for the design of the ATOCA logo. This project was undertaken with the financial support of the Canadian Space Agency (CSA-ASC) and the \textit{Fonds de Recherche du Qu\'ebec en Nature et Technologies} (FRQNT). We would also like to thank the Space Telescope Science Institute (STScI) for their trust and help during the implementation process. A.D.B., C.P. and S.P. wants to thank the Technologies for Exo-Planetary Science (TEPS) CREATE program, without whom this research would not be possible. The authors also acknowledge financial and social support of the Institute for Research on Exoplanets (iREx) and the University of Montreal.
MR would like to acknowledge funding from FRQNT, as well as the National Sciences and Engineering Research Council of Canada (NSERC). 
C.P. acknowledges financial support by the NSERC Vanier Scholarship.
D.J. is supported by NRC Canada and by an NSERC Discovery Grant.
Support for J.D.T. was provided by NASA through the NASA Hubble Fellowship grant \# HST-HF2-51495.001-A awarded by the Space Telescope Science Institute, which is operated by the Association of Universities for Research in Astronomy, Incorporated, under NASA contract NAS5-26555.

\software{\texttt{WebbPSF} \citep{webbpsf}, JWST Data Management System \url{https://jwst-pipeline.readthedocs.io/en/latest/index.html}, \texttt{scipy} \citep{scipy.2020}, \texttt{ipython}  \citep{ipython.2007}, \texttt{matplotlib} \citep{matplotlib.2007}, \texttt{numpy} \citep{numpy2020}.}

\bibliographystyle{apj}
\bibliography{bibliography}

\begin{thebibliography}{}
\expandafter\ifx\csname natexlab\endcsname\relax\def\natexlab#1{#1}\fi

\bibitem[{{Batalha} {et~al.}(2017){Batalha}, {Bean}, {Stevenson}, {Alam},
  {Batalha}, {Benneke}, {Berta-Thompson}, {Blecic}, {Bruno}, {Carter},
  {Chapman}, {Crossfield}, {Crouzet}, {Decin}, {Demory}, {Desert}, {Dragomir},
  {Fortney}, {Fraine}, {Gao}, {Garcia Munoz}, {Gibson}, {Goyal}, {Harrington},
  {Heng}, {Hu}, {Kempton}, {Kendrew}, {Kilpatrick}, {Knutson}, {Kreidberg},
  {Krick}, {Lagage}, {Lendl}, {Line}, {Lopez-Morales}, {Louden}, {Madhusudhan},
  {Mandell}, {Mansfield}, {May}, {Mikal-Evans}, {Morello}, {Morley}, {Moses},
  {Nikolov}, {Parmentier}, {Redfield}, {Roberts}, {Schlawin}, {Showman},
  {Sing}, {Spake}, {Swain}, {Todorov}, {Tsiaras}, {Venot}, {Waalkes},
  {Wakeford}, {Wheatley}, \& {Zellem}}]{1366jwst}
{Batalha}, N., {Bean}, J.~L., {Stevenson}, K.~B., {et~al.} 2017, {The
  Transiting Exoplanet Community Early Release Science Program}, JWST Proposal
  ID 1366. Cycle 0 Early Release Science

\bibitem[{{Batalha} \& {Line}(2017)}]{batalha.2017}
{Batalha}, N.~E., \& {Line}, M.~R. 2017, \aj, 153, 151

\bibitem[{{Bean} {et~al.}(2018){Bean}, {Stevenson}, {Batalha},
  {Berta-Thompson}, {Kreidberg}, {Crouzet}, {Benneke}, {Line}, {Sing},
  {Wakeford}, {Knutson}, {Kempton}, {D{\'e}sert}, {Crossfield}, {Batalha}, {de
  Wit}, {Parmentier}, {Harrington}, {Moses}, {Lopez-Morales}, {Alam}, {Blecic},
  {Bruno}, {Carter}, {Chapman}, {Decin}, {Dragomir}, {Evans}, {Fortney},
  {Fraine}, {Gao}, {Garc{\'\i}a Mu{\~n}oz}, {Gibson}, {Goyal}, {Heng}, {Hu},
  {Kendrew}, {Kilpatrick}, {Krick}, {Lagage}, {Lendl}, {Louden}, {Madhusudhan},
  {Mandell}, {Mansfield}, {May}, {Morello}, {Morley}, {Nikolov}, {Redfield},
  {Roberts}, {Schlawin}, {Spake}, {Todorov}, {Tsiaras}, {Venot}, {Waalkes},
  {Wheatley}, {Zellem}, {Angerhausen}, {Barrado}, {Carone}, {Casewell},
  {Cubillos}, {Damiano}, {de Val-Borro}, {Drummond}, {Edwards}, {Endl},
  {Espinoza}, {France}, {Gizis}, {Greene}, {Henning}, {Hong}, {Ingalls}, {Iro},
  {Irwin}, {Kataria}, {Lahuis}, {Leconte}, {Lillo-Box}, {Lines}, {Lothringer},
  {Mancini}, {Marchis}, {Mayne}, {Palle}, {Rauscher}, {Roudier}, {Shkolnik},
  {Southworth}, {Swain}, {Taylor}, {Teske}, {Tinetti}, {Tremblin}, {Tucker},
  {van Boekel}, {Waldmann}, {Weaver}, \& {Zingales}}]{bean.2018}
{Bean}, J.~L., {Stevenson}, K.~B., {Batalha}, N.~M., {et~al.} 2018, \pasp, 130,
  114402

\bibitem[{Benneke(2015)}]{benneke.2015}
Benneke, B. 2015, arXiv:1504.07655 [astro-ph], arXiv: 1504.07655

\bibitem[{Benneke \& Seager(2012)}]{benneke_atmospheric_2012}
Benneke, B., \& Seager, S. 2012, ApJ, 753, 100

\bibitem[{Benneke \& Seager(2013)}]{benneke_how_2013}
---. 2013, ApJ, 778, 153

\bibitem[{Berta {et~al.}(2011)Berta, Charbonneau, Bean, Irwin, Burke, Désert,
  Nutzman, \& Falco}]{berta.2011}
Berta, Z.~K., Charbonneau, D., Bean, J., {et~al.} 2011, The Astrophysical
  Journal, 736, 12

\bibitem[{Bolton \& Schlegel(2010)}]{bolton.2010}
Bolton, A.~S., \& Schlegel, D.~J. 2010, Publications of the Astronomical
  Society of the Pacific, 122, 248, aDS Bibcode: 2010PASP..122..248B

\bibitem[{{de Boer} \& {Snijders}(1981)}]{deboer.1981}
{de Boer}, K.~S., \& {Snijders}, M.~A.~J. 1981, IUE ESA Newsletter, 14, 154

\bibitem[{Deming {et~al.}(2013)Deming, Wilkins, McCullough, Burrows, Fortney,
  Agol, Dobbs-Dixon, Madhusudhan, Crouzet, Desert, Gilliland, Haynes, Knutson,
  Line, Magic, Mandell, Ranjan, Charbonneau, Clampin, Seager, \&
  Showman}]{deming.2013}
Deming, D., Wilkins, A., McCullough, P., {et~al.} 2013, The Astrophysical
  Journal, 774, 95, aDS Bibcode: 2013ApJ...774...95D

\bibitem[{{Diamond-Lowe} {et~al.}(2018){Diamond-Lowe}, {Berta-Thompson},
  {Charbonneau}, \& {Kempton}}]{diamond.2018}
{Diamond-Lowe}, H., {Berta-Thompson}, Z., {Charbonneau}, D., \& {Kempton}, E.
  M.~R. 2018, \aj, 156, 42

\bibitem[{{Espinoza} {et~al.}(2021{\natexlab{a}}){Espinoza}, {Loic},
  {Goudfrooij}, {Martel}, {Vila}, {Volk}, \& Julia}]{1541jwst}
{Espinoza}, N., {Loic}, A., {Goudfrooij}, P., {et~al.} 2021{\natexlab{a}},
  {NIRISS Sensitivity and Stability for Transiting Exoplanet Observations},
  JWST Proposal. Cycle 0

\bibitem[{{Espinoza} {et~al.}(2021{\natexlab{b}}){Espinoza}, {Baeyens},
  {Bello-Arufe}, {Buchhave}, {Burgasser}, {Carone}, {Diamond-Lowe}, {Fisher},
  {Gibson}, {Guzman Mesa}, {Heng}, {Hoeijmakers}, {Hooton}, {Kitzmann},
  {Kozakis}, {Lopez-Morales}, {Mendonca}, {Molliere}, {Morris}, \&
  {Rathcke}}]{2113jwst}
{Espinoza}, N., {Baeyens}, R., {Bello-Arufe}, A., {et~al.} 2021{\natexlab{b}},
  {Exploring the morning and evening limbs of a transiting exoplanet}, JWST
  Proposal. Cycle 1, ID. \#2113

\bibitem[{Evans {et~al.}(2017)Evans, Sing, Kataria, Goyal, Nikolov, Wakeford,
  Deming, Marley, Amundsen, Ballester, Barstow, Ben-Jaffel, Bourrier, Buchhave,
  Cohen, Ehrenreich, García~Muñoz, Henry, Knutson, Lavvas, Etangs, Lewis,
  López-Morales, Mandell, Sanz-Forcada, Tremblin, \& Lupu}]{evans.2017}
Evans, T.~M., Sing, D.~K., Kataria, T., {et~al.} 2017, Nature, 548, 58

\bibitem[{Fitzpatrick(2013)}]{canada}
Fitzpatrick, S. 2013, Cranberry

\bibitem[{{Genest} {et~al.}(2022){Genest}, {Lafreni{\`e}re}, {Boucher},
  {Darveau-Bernier}, {Doyon}, {Artigau}, \& {Cook}}]{genest.2022}
{Genest}, F., {Lafreni{\`e}re}, D., {Boucher}, A., {et~al.} 2022, arXiv
  e-prints, arXiv:2205.09859

\bibitem[{Golub {et~al.}(1979)Golub, Heath, \& Wahba}]{golub.1979}
Golub, G.~H., Heath, M., \& Wahba, G. 1979, Technometrics, 21, 215

\bibitem[{Green \& Silverman(1993)}]{green1993}
Green, P.~J., \& Silverman, B.~W. 1993, Nonparametric regression and
  generalized linear models: a roughness penalty approach (Crc Press)

\bibitem[{{Greene} {et~al.}(2016){Greene}, {Line}, {Montero}, {Fortney},
  {Lustig-Yaeger}, \& {Luther}}]{greene.2016}
{Greene}, T.~P., {Line}, M.~R., {Montero}, C., {et~al.} 2016, \apj, 817, 17

\bibitem[{Hansen(1992)}]{hansen1992}
Hansen, P.~C. 1992, SIAM review, 34, 561

\bibitem[{Hansen \& O’Leary(1993)}]{hansen1993}
Hansen, P.~C., \& O’Leary, D.~P. 1993, SIAM journal on scientific computing,
  14, 1487

\bibitem[{Harris {et~al.}(2020)Harris, Millman, van~der Walt, Gommers,
  Virtanen, Cournapeau, Wieser, Taylor, Berg, Smith, Kern, Picus, Hoyer, van
  Kerkwijk, Brett, Haldane, del R{\'{i}}o, Wiebe, Peterson,
  G{\'{e}}rard-Marchant, Sheppard, Reddy, Weckesser, Abbasi, Gohlke, \&
  Oliphant}]{numpy2020}
Harris, C.~R., Millman, K.~J., van~der Walt, S.~J., {et~al.} 2020, Nature, 585,
  357

\bibitem[{Hastie \& Tibshirani(1990)}]{hastie.1990}
Hastie, T.~J., \& Tibshirani, R. 1990, 35

\bibitem[{Horel(1962)}]{horel.1962}
Horel, A. 1962, Chem. Eng. Progress., 58, 54

\bibitem[{{Horne}(1986)}]{horne.1986}
{Horne}, K. 1986, \pasp, 98, 609

\bibitem[{Hunter(2007)}]{matplotlib.2007}
Hunter, J.~D. 2007, Computing in Science \& Engineering, 9, 90

\bibitem[{{Husser} {et~al.}(2013){Husser}, {Wende-von Berg}, {Dreizler},
  {Homeier}, {Reiners}, {Barman}, \& {Hauschildt}}]{husser.2013}
{Husser}, T.~O., {Wende-von Berg}, S., {Dreizler}, S., {et~al.} 2013, \aap,
  553, A6

\bibitem[{{Hynes}(2002)}]{hynes.2002}
{Hynes}, R.~I. 2002, \aap, 382, 752

\bibitem[{Jordán {et~al.}(2013)Jordán, Espinoza, Rabus, Eyheramendy, Sing,
  Désert, Bakos, Fortney, López-Morales, Maxted, Triaud, \&
  Szentgyorgyi}]{jordan.2013}
Jordán, A., Espinoza, N., Rabus, M., {et~al.} 2013, ApJ, 778, 184

\bibitem[{{Kempton} {et~al.}(2021){Kempton}, {Bean}, {Deming}, {Fu}, {Gao},
  {Ih}, {Malik}, {May}, \& {Rogers}}]{1935jwst}
{Kempton}, E. M.~R., {Bean}, J.~L., {Deming}, D., {et~al.} 2021, {Unshrouding
  the Sub-Neptune Population: The Case of TOI-421b}, JWST Proposal. Cycle 1,
  ID. \#1935

\bibitem[{Khmil \& Surdej(2002)}]{khmil.2002}
Khmil, S.~V., \& Surdej, J. 2002, Astronomy \& Astrophysics, 387, 347

\bibitem[{Kreidberg {et~al.}(2014)Kreidberg, Bean, Désert, Benneke, Deming,
  Stevenson, Seager, Berta-Thompson, Seifahrt, \& Homeier}]{kreidberg.2014}
Kreidberg, L., Bean, J.~L., Désert, J.-M., {et~al.} 2014, Nature, 505, 69,
  arXiv:1401.0022 [astro-ph]

\bibitem[{{Kunasz} {et~al.}(1973){Kunasz}, {Jefferies}, \&
  {White}}]{kunasz.1973}
{Kunasz}, C.~V., {Jefferies}, J.~T., \& {White}, O.~R. 1973, \aap, 28, 15

\bibitem[{{Lafreniere}(2017)}]{1201jwst}
{Lafreniere}, D. 2017, {NIRISS Exploration of the Atmospheric diversity of
  Transiting exoplanets (NEAT)}, JWST Proposal. Cycle 1, ID. \#1201

\bibitem[{{Li} {et~al.}(2015){Li}, {Zhang}, \& {Bai}}]{lamost_tikhonov.2015}
{Li}, G., {Zhang}, H., \& {Bai}, Z. 2015, arXiv e-prints, arXiv:1506.06188

\bibitem[{{Lim} {et~al.}(2021){Lim}, {Albert}, {Artigau}, {Benneke}, {Cowan},
  {Doyon}, \& {Lafreniere}}]{2589jwst}
{Lim}, O., {Albert}, L., {Artigau}, E., {et~al.} 2021, {Atmospheric
  reconnaissance of the TRAPPIST-1 planets}, JWST Proposal. Cycle 1, ID. \#2589

\bibitem[{{Louie} {et~al.}(2018){Louie}, {Deming}, {Albert}, {Bouma}, {Bean},
  \& {Lopez-Morales}}]{louie.2018}
{Louie}, D.~R., {Deming}, D., {Albert}, L., {et~al.} 2018, \pasp, 130, 044401

\bibitem[{Lucy \& Walsh(2003)}]{lucy.2003}
Lucy, L.~B., \& Walsh, J.~R. 2003, The Astronomical Journal, 125, 2266, aDS
  Bibcode: 2003AJ....125.2266L

\bibitem[{{Madhusudhan} {et~al.}(2021){Madhusudhan}, {Constantinou}, {Moses},
  {Piette}, \& {Sarkar}}]{2722jwst}
{Madhusudhan}, N., {Constantinou}, S., {Moses}, J.~I., {Piette}, A., \&
  {Sarkar}, S. 2021, {Chemical Disequilibrium in a Temperate sub-Neptune}, JWST
  Proposal. Cycle 1, ID. \#2722

\bibitem[{Marsh(1989)}]{marsh.1989}
Marsh, T.~R. 1989, PASP, 101, 1032

\bibitem[{{Martel} {et~al.}(2020){Martel}, {Filippazzo}, {Volk}, {Loic},
  {Goudfrooij}, {Vila}, \& Julia}]{1091jwst}
{Martel}, A., {Filippazzo}, J., {Volk}, K., {et~al.} 2020, {NIRISS GR700XD Flux
  Calibration}, JWST Proposal. Cycle 0

\bibitem[{{Mayo} {et~al.}(2021){Mayo}, {Dressing}, {Fortenbach}, {Giacalone},
  {Harada}, \& {Turtelboom}}]{2062jwst}
{Mayo}, A., {Dressing}, C., {Fortenbach}, C., {et~al.} 2021, {Transmission
  Spectroscopy of the Super-Neptune WASP-166b}, JWST Proposal. Cycle 1, ID.
  \#2062

\bibitem[{Mikal-Evans {et~al.}(2021)Mikal-Evans, Crossfield, Benneke,
  Kreidberg, Moses, Morley, Thorngren, Mollière, Hardegree-Ullman, Brewer,
  Christiansen, Ciardi, Dragomir, Dressing, Fortney, Gorjian, Greene, Hirsch,
  Howard, Howell, Isaacson, Kosiarek, Krick, Livingston, Lothringer, Morales,
  Petigura, Schlieder, \& Werner}]{mikal-evans.2021}
Mikal-Evans, T., Crossfield, I. J.~M., Benneke, B., {et~al.} 2021, The
  Astronomical Journal, 161, 18, aDS Bibcode: 2021AJ....161...18M

\bibitem[{Min {et~al.}(2020)Min, Guang-wei, Ke, Fu-qing, Haerken, \&
  Yong-heng}]{min.2020}
Min, L., Guang-wei, L., Ke, L., {et~al.} 2020, Chinese Astronomy and
  Astrophysics, 44, 399

\bibitem[{P\'erez \& Granger(2007)}]{ipython.2007}
P\'erez, F., \& Granger, B.~E. 2007, Computing in Science and Engineering, 9,
  21

\bibitem[{{Perrin} {et~al.}(2015){Perrin}, {Long}, {Sivaramakrishnan},
  {Lajoie}, {Elliot}, {Pueyo}, \& {Albert}}]{webbpsf}
{Perrin}, M.~D., {Long}, J., {Sivaramakrishnan}, A., {et~al.} 2015, {WebbPSF:
  James Webb Space Telescope PSF Simulation Tool}, Astrophysics Source Code
  Library, record ascl:1504.007, ascl:1504.007

\bibitem[{Phillips(1962)}]{phillips.1962}
Phillips, D.~L. 1962, J. ACM, 9, 84–97

\bibitem[{Piskunov {et~al.}(2021)Piskunov, Wehrhahn, \&
  Marquart}]{piskunov_optimal_2021}
Piskunov, N., Wehrhahn, A., \& Marquart, T. 2021, Astronomy and Astrophysics,
  646, A32

\bibitem[{Puetter {et~al.}(2005)Puetter, Gosnell, \&
  Yahil}]{puetter_digital_2005}
Puetter, R., Gosnell, T., \& Yahil, A. 2005, Annual Review of Astronomy and
  Astrophysics, 43, 139

\bibitem[{{Rackham} {et~al.}(2018){Rackham}, {Apai}, \&
  {Giampapa}}]{rackham.2018}
{Rackham}, B.~V., {Apai}, D., \& {Giampapa}, M.~S. 2018, \apj, 853, 122

\bibitem[{Robertson(1986)}]{robertson.1986}
Robertson, J.~G. 1986, Publications of the Astronomical Society of the Pacific,
  98, 1220, aDS Bibcode: 1986PASP...98.1220R

\bibitem[{{Schlawin} {et~al.}(2018){Schlawin}, {Greene}, {Line}, {Fortney}, \&
  {Rieke}}]{schlawin.2018}
{Schlawin}, E., {Greene}, T.~P., {Line}, M., {Fortney}, J.~J., \& {Rieke}, M.
  2018, \aj, 156, 40

\bibitem[{Sing {et~al.}(2015)Sing, Wakeford, Showman, Nikolov, Fortney,
  Burrows, Ballester, Deming, Aigrain, Désert, Gibson, Henry, Knutson,
  Lecavelier~des Etangs, Pont, Vidal-Madjar, Williamson, \& Wilson}]{sing.2015}
Sing, D.~K., Wakeford, H.~R., Showman, A.~P., {et~al.} 2015, Monthly Notices of
  the Royal Astronomical Society, 446, 2428

\bibitem[{{Spake} {et~al.}(2021){Spake}, {Rustamkulov}, {Benneke}, {Fortney},
  {Fu}, {Lothringer}, {Moran}, {Piaulet}, {Schlaufman}, {Sing}, \&
  {Thorngren}}]{2594jwst}
{Spake}, J., {Rustamkulov}, Z., {Benneke}, B., {et~al.} 2021, {The twin
  paradox: assessing planetary radius evolution with a CH4 thermometer}, JWST
  Proposal. Cycle 1, ID. \#2594

\bibitem[{Stevenson {et~al.}(2014)Stevenson, Bean, Seifahrt, Désert,
  Madhusudhan, Bergmann, Kreidberg, \& Homeier}]{stevenson.2014}
Stevenson, K.~B., Bean, J.~L., Seifahrt, A., {et~al.} 2014, The Astronomical
  Journal, 147, 161

\bibitem[{{Stevenson} \& {Fowler}(2019)}]{stevenson.2019}
{Stevenson}, K.~B., \& {Fowler}, J. 2019, {Analyzing Eight Years of Transiting
  Exoplanet Observations Using WFC3's Spatial Scan Monitor}, Instrument Science
  Report WFC3 2019-12, 16 pages, arXiv:1910.02073

\bibitem[{{Thompson}(1990)}]{thompson.1990}
{Thompson}, A.~M. 1990, \aap, 240, 209

\bibitem[{{Tikhonov}(1963)}]{tikhonov.1963}
{Tikhonov}, A.~N. 1963, in Doklady Akademii Nauk, Vol. 151, Russian Academy of
  Sciences, 501--504

\bibitem[{{TLFQ}(1998)}]{dhfq}
{TLFQ}. 1998, in Dictionnaire historique du fran\c cais qu\'eb\'ecois
  (Qu\'ebec: Presses de l'Universit\'e Laval)

\bibitem[{Virtanen {et~al.}(2020)Virtanen, Gommers, Oliphant, Haberland, Reddy,
  Cournapeau, Burovski, Peterson, Weckesser, Bright, {van der Walt}, Brett,
  Wilson, Millman, Mayorov, Nelson, Jones, Kern, Larson, Carey, Polat, Feng,
  Moore, {VanderPlas}, Laxalde, Perktold, Cimrman, Henriksen, Quintero, Harris,
  Archibald, Ribeiro, Pedregosa, {van Mulbregt}, \& {SciPy 1.0
  Contributors}}]{scipy.2020}
Virtanen, P., Gommers, R., Oliphant, T.~E., {et~al.} 2020, Nature Methods, 17,
  261

\bibitem[{Wahba(1977)}]{wahba.1977}
Wahba, G. 1977, SIAM journal on numerical analysis, 14, 651

\bibitem[{Wakeford {et~al.}(2013)Wakeford, Sing, Deming, Gibson, Fortney,
  Burrows, Ballester, Nikolov, Aigrain, Henry, Knutson, Lecavelier~des Etangs,
  Pont, Showman, Vidal-Madjar, \& Zahnle}]{wakeford.2013}
Wakeford, H.~R., Sing, D.~K., Deming, D., {et~al.} 2013, Monthly Notices of the
  Royal Astronomical Society, 435, 3481

\end{thebibliography}



\appendix

\section{Minimization of the $\chi^2$}
\label{ssec:chi2}

From equation \ref{eq:chi2}, we can compute the derivative of the $\chi^2$:
\begin{equation}
\begin{aligned}
    \frac{d \chi^2}{d f_k} 
        & = \frac{d}{d f_k}
            \sum_i \frac{1}{2}\left(\frac{D_i-M_i}{\sigma_i}\right)^2 \\
        & = \sum_i -\left(\frac{D_i-M_i}{\sigma_i}\right)
            \frac{d}{d f_k} \frac{M_i}{\sigma_i} \\
        & = \sum_i -\left(\frac{D_i-\sum_{k\prime}B_{ik\prime}f_{k\prime}}{\sigma_i}\right)
            \frac{d}{d f_k} \frac{1}{\sigma_i} \sum_{k\prime\prime}B_{ik\prime\prime}f_{k\prime\prime} \, .
\end{aligned}
\end{equation}
However, since the $B_{ik\prime\prime}$ are simple coefficients, we have
\begin{equation}
        \frac{d}{d f_k}  \sum_{k\prime\prime}B_{ik\prime\prime}f_{k\prime\prime} 
        = \sum_{k\prime\prime}B_{ik\prime\prime}\frac{d f_{k\prime\prime}}{d f_k} 
        = \sum_{k\prime\prime}B_{ik\prime\prime} \delta_{kk\prime\prime}
        = B_{ik} \, ,
\end{equation}
to finally end up with
\begin{equation}
     \frac{d \chi^2}{d f_k} = \sum_i -\left(\frac{D_i-\sum_{k\prime}B_{ik\prime}f_{k\prime}}{\sigma_i}\right) \frac{B_{ik}}{\sigma_i} \, .
\end{equation}
By setting this equation equal to zero, we have
\begin{equation}
\begin{aligned}
    0 = \sum_i -\frac{D_i}{\sigma_i} \frac{B_{ik}}{\sigma_i}
    + \sum_i \sum_{k\prime}\frac{B_{ik}}{\sigma_i}\frac{B_{ik\prime}}{\sigma_i}f_{k\prime} \, ,
\end{aligned}
\end{equation}
or
\begin{equation}\label{eq:lin_sys}
    \sum_i \frac{D_i}{\sigma_i} \frac{B_{ik}}{\sigma_i}
    = \sum_i \sum_{k'} \frac{B_{ik}}{\sigma_i} \frac{B_{ik'}}{\sigma_i} f_{k'}\, .
\end{equation}

\section{Trapezoidal integration on a grid} \label{sec:trpz}
Consider the following integral:
\begin{equation}
M_{i} = \sum_n \int_{\lambda_{ni}^-}^{\lambda_{ni}^+} P_{ni}T_n(\lambda)\tilde{f}_n(\lambda)\lambda d\lambda
= \sum_n \int_{\lambda_{ni}^-}^{\lambda_{ni}^+} \tilde{g}_{ni}(\lambda) d\lambda,
\end{equation}
with $\tilde{g}$ being the integrand. We can approximate the integral in a discrete form with the trapezoidal method,
\begin{equation}\label{eq:model}
M_{i} \simeq \sum_n \left(\frac{\tilde{g}_{ni}(\lambda_{ni}^-) + \tilde{g}_{niL_{ni}}}{2}
(\lambda_{L_{ni}}-\lambda_{ni}^-)	+ \sum_{\tilde{k}=L_{ni}+1}^{H_{ni}-1} \frac{\tilde{g}_{ni\tilde{k}} + \tilde{g}_{ni(\tilde{k}+1)}}{2}
\Delta \lambda_{\tilde{k}}	+ \frac{\tilde{g}_{niH_{ni}} + \tilde{g}_{ni}(\lambda_{ni}^+)}{2} (\lambda_{ni}^+ - \lambda_{H_{ni}}) \right) \, ,
\end{equation}
where $L_{ni}$ and $H_{ni}$ are respectively the lowest and highest $\tilde{k}$ located into the pixel, so that $L_{ni}$ is the smallest $\tilde{k}$ for which $\lambda_{ni}^-  \leq \lambda_{\tilde{k}}$ and $H_{ni}$ is the biggest $\tilde{k}$ for which $\lambda_{\tilde{k}} \leq \lambda_{ni}^+$. Now, we need to express $\tilde{g}_{ni}(\lambda_{ni}^-)$ and $\tilde{g}_{ni}(\lambda_{ni}^+)$ with the $\tilde{g}_{ni\tilde{k}}$. To do so, we can use linear interpolation to write
\begin{equation}\label{eq:g_borders}
\begin{aligned}
\tilde{g}(\lambda_i^-) = \frac{\lambda_{L_{ni}} - \lambda_i^-}{\Delta \lambda_{L_{ni}-1}} \tilde{g}_{L_{ni}-1} + \frac{\lambda_i^- - \lambda_{L_{ni}-1}}{\Delta \lambda_{L_{ni}-1}} \tilde{g}_{L_{ni}} \,\\
\tilde{g}(\lambda_i^+) = \frac{\lambda_{H_{ni}+1} - \lambda_i^+}{\Delta \lambda_{H_{ni}}} \tilde{g}_{H_{ni}} + \frac{\lambda_i^+ - \lambda_{H_{ni}}}{\Delta \lambda_{H_{ni}}} \tilde{g}_{H_{ni}+1}.
\end{aligned}
\end{equation}

Thus, this approximation will be permitted for pixels where $\lambda_i^- \geq \lambda_{\tilde{k}=1}$ and $\lambda_i^+ \leq \lambda_{\tilde{k}=N_{\tilde{k}}}$. We can now substitute equation \ref{eq:g_borders} into equation \ref{eq:model} and, after some algebra, we get
\begin{equation}
\begin{aligned}
M_{i} = \sum_n \frac{1}{2}\Biggl( & \frac{(\lambda_{L_{ni}}-\lambda_i^-)^2}{ \Delta \lambda_{L_{ni} - 1}} \tilde{g}_{L_{ni}-1} \\
 &+ \left( \lambda_{L_{ni}+1} -  \lambda_i^- + \frac{(\lambda_i^- - \lambda_{L_{ni}-1})(\lambda_{L_{ni}}-\lambda_i^-)}{\Delta \lambda_{L_{ni} - 1}}\right) \tilde{g}_{L_{ni}} \\
 &+ \sum_{\tilde{k}=L_{ni} + 1}^{H_{ni}-1} \left(\Delta\lambda_{\tilde{k}}+\Delta \lambda_{\tilde{k}-1}\right) \tilde{g}_{\tilde{k}} \\
 &+ \left(\lambda_i^+ - \lambda_{H_{ni}-1} + \frac{(\lambda_i^+ - \lambda_{H_{ni}})(\lambda_{H_{ni}+1}-\lambda_i^+)}{\Delta\lambda_{H_{ni}}} \right) \tilde{g}_{H_{ni}} \\
 &+ \frac{(\lambda_i^+ - \lambda_{H_{ni}})^2}{\Delta \lambda_{H_{ni}}} \tilde{g}_{H_{ni}+1} \Biggr) \, .
\end{aligned}
\end{equation}
This has the form needed to compute the $w_{ni\tilde{k}}$ in the equation
\begin{equation}
    M_i = \sum_n \sum_{\tilde{k}=L_{ni}-1}^{H_{ni}+1} w_{ni\tilde{k}} \tilde{g}_{\tilde{k}} \, .
\end{equation}





\section{Comparison to optimal extraction method} \label{sec:comparison}


As highlighted in \cite{horne.1986}, the optimal extraction can be seen as the best correspondence (in the sense of a $\chi^2$ minimization) between the sky-subtracted data and a model of the detector. The latter is a function of the incident flux expressed as a scaling of the spatial profile. This is also the main idea of ATOCA and its relation with the optimal extraction method in the absence of contamination is demonstrated in the following lines.

Let's consider the situation where there is no tilt in the wavelength solution and only one order is present so there is no contamination. Then, the integration weights, $w_{i\tilde{k}}$ (see \ref{eq:pix_model_num}),  will only depend on the spectral axis. Moreover, by invoking the mean value theorem, the integral from equation \ref{eq:pix_model} becomes
\begin{equation}
    M_i = (\lambda_i^+ - \lambda_i^-) \cdot P_i \cdot T(\lambda_i^*) \cdot \tilde{f}(\lambda_i^*) \cdot \lambda_i^*
\end{equation}
where $\lambda_i^* \in \, [\lambda_i^-, \lambda_i^+] $. Thus, by choosing the integration grid to be the $\lambda_i^*$, the factors from equation \ref{eq:pix_model_num} are given by $w_{ik}=(\lambda_i^+ - \lambda_i^-)$, $a_{k} = \lambda_i^* \cdot  T(\lambda_i^*)$ and $\tilde{f}_k = \tilde{f}(\lambda_i^*)$. Furthermore, since the wavelength solution is fixed along the spatial axis, the index $i$ can be removed from the notation, except for $P_i$. This leads to the model of a pixel given exactly by
\begin{equation}
    M_i = P_i \, w_k \, a_k \, \tilde{f_k}.
\end{equation}

Finally, after $\chi^2$ minimization with the procedure described in Section \ref{ssec:chi2}, the best solution for $\tilde{f_k}$ is given by a linear equation where each row is given by
\begin{equation}
    \sum_i \frac{D_i\cdot P_i}{\sigma_i^2} = \sum_i \frac{P_i^2}{\sigma_i^2} w_k \, a_k \, \tilde{f_k} 
    = \sum_i \frac{P_i^2}{\sigma_i^2} S_k \, ,
\end{equation}
where $S_k = w_k \, a_k \, \tilde{f_k}$ is the integrated signal over a pixel, which corresponds to the results of an optimal extraction. Each row of this linear system is independent, so the solution is directly obtained by re-arranging the terms, resulting in the equation
\begin{equation}
    S_k =  \sum_i \frac{D_i\cdot P_i / \sigma_i^2}{\sum_i P_i^2/\sigma_i^2},
\end{equation}
which is the exact definition of the optimal extraction.

This means that the determination of the spatial profile can be inspired by the optimal extraction. Moreover, considering that the product of ATOCA is similar to one of an optimal extraction method, it thus constitutes an alternative to optimal extraction that can account for a distorted wavelength solution. However, in this scenario, the integral will not be solved exactly for each pixel, leaving the extraction subject to numerical integration errors. Yet, these errors can be mitigated with a good choice of integration method or an oversampled grid.




\end{document}